\documentclass[journal,a4paper]{IEEEtran}
\usepackage[font=normal,labelfont=bf]{caption} 
\usepackage{cite}
\usepackage{amsmath,amssymb,amsfonts}
\usepackage{algorithmic}
\usepackage{graphicx}
\usepackage{subfigure}
\usepackage{textcomp}
\usepackage{capt-of}
\usepackage{multirow}
\usepackage{textcomp}
\usepackage{siunitx} 
\usepackage[normalem]{ulem}
\usepackage{xcolor}
\usepackage{multirow}
\usepackage{bm}
\usepackage{cleveref}
\usepackage{booktabs}
\usepackage[utf8]{inputenc}
\usepackage{comment}
\usepackage{titlesec}
\usepackage{xurl}
\usepackage{threeparttable}
\usepackage{soul}
\usepackage{array}
\usepackage{flushend}

\titleformat{\subparagraph}[runin]{\normalfont\normalsize\bfseries}{\thesubparagraph}{1em}{}
\titlespacing*{\subparagraph}{0pt}{3ex}{1ex}

\renewcommand{\j}{\mathrm{j}}
\newcommand{\CI}{\mathrm{CI}}
\newcommand{\FI}{\mathrm{FI}}
\newcommand{\ABG}{\mathrm{ABG}}
\newcommand{\R}{\mathrm{R}}
\newcommand{\T}{\mathrm{T}}

\newcommand{\PL}{\mathrm{PL}}
\newcommand{\DS}{\mathrm{DS}}

\newcommand{\ASD}{\mathrm{ASD}}
\newcommand{\ASA}{\mathrm{ASA}}

\newcommand{\ThreeD}{\mathrm{3D}}
\newcommand{\GHz}{\mathrm{GHz}}
\newcommand{\dB}{\mathrm{dB}}

\newcommand{\Vect}[1]{\boldsymbol{#1}}

\newcommand{\NA}{---}

\newif\ifMarking
\Markingtrue
\ifMarking
    
\else
    
\fi

\def\BibTeX{{\rm B\kern-.05em{\sc i\kern-.025em b}\kern-.08em
    T\kern-.1667em\lower.7ex\hbox{E}\kern-.125emX}}

\begin{document}

\title{Urban Macro/Microcellular Channel Characterization at 4.85~GHz With Literature-Referenced Upper FR1-to-FR3 Cross-Band Analysis}

\author{Inocent Calist, \IEEEmembership{Graduate Student Member, IEEE}, Minseok Kim, \IEEEmembership{Senior Member, IEEE}
\thanks{The authors are with the Graduate School of Science and Technology, Niigata University, 8050 Ikarashi 2-no-cho, Nishi-ku, Niigata, 950-2181, Japan.}
\thanks{This research has been conducted under the contract (\#JPJ000254 and \#JPMI240410003) made with the Ministry of Internal Affairs and Communications of Japan.}
\thanks{Corresponding author: Minseok Kim (mskim@eng.niigata-u.ac.jp)}}

\maketitle

\begin{abstract}
The transition from 5G to 6G requires frequency-dependent, physically consistent radio channel models across the Upper FR1/FR3 transition region, particularly in the under-explored $4$--$8$~GHz region targeted in the current WRC-$27$ studies, where outdoor urban channel measurements and characterizations remain scarce. This paper presents a $4.85$~GHz measurement-anchored study of urban channels and a literature-referenced cross-band analysis. Double-directional measurements were conducted at $4.85$~GHz in urban macrocell (UMa) and urban microcell (UMi) routes in Yokohama, Japan, from which path loss, delay spread (DS), azimuth spread of arrival/departure (ASA/ASD), $K$-factor, and route-dependent spatial-consistency statistics were extracted. To align these results in a broader cross-band context, the measured $4.85$~GHz large-scale parameter (LSP) means were combined with scenario-matched literature anchors to derive log-log trends for DS, ASA, and ASD over an approximately $4$--$28$~GHz range around the $7.125$~GHz Upper FR1/FR3 cross-band boundary. The resulting trends were compared with 3GPP UMa/UMi reference parameterizations over the same interval, and the sensitivity of the UMi DS fit was examined via leave-one-out analysis. Because the cross-band analysis still relies on a single in-house measurement band alongside heterogeneous anchors from different campaigns, it is presented as measurement-informed and indicative rather than as a definitive multi-band model. The paper therefore contributes both a detailed, parameterized $4.85$~GHz urban measurement reference and a bounded literature-referenced view of channel behavior near the Upper FR1/FR3 transition.
\end{abstract}


\section{Introduction}
\IEEEPARstart {T}{he} evolution from fifth-generation (5G) toward sixth-generation (6G) networks has increased interest in the upper mid-band spectrum, particularly the $7.125$--$24.25$~GHz range above the traditional sub-6~GHz bands  \cite{Samsung_Research}. This spectrum promises to support higher data rates, ultra-low latency, and enhanced support for emerging capabilities such as integrated sensing and communications (ISAC). This will address the growing demands for spectral efficiency, wide-area connectivity, and device-intensive applications, including augmented reality, autonomous driving, and real-time analytics \cite{[Wang]B5GTrends, WRC_ITU}. A key challenge in this evolution is the availability of radio channel models that remain frequency-dependent and physically consistent across the transition from Frequency Range~1 (FR1, $\leq 7.125$~GHz) to Frequency Range~3 (FR3, $7.125$--$24.25$~GHz) \cite{WRC-23, ITU-R_P.1411, Samsung_Research, NR_3GPP}. This need is especially acute in the under-explored $4$--$8$~GHz region, which has become increasingly relevant in ongoing WRC-$27$ and IMT-$2030$ discussions \cite{WRC_27, Comprehensive_FR1C_FR3_NYU, ITU-R_2160}.

In this paper, the term Upper FR1/FR3 transition denotes the frequency region around the $7.125$~GHz boundary between FR1 and FR3. It does not imply that FR2 or the broader mmWave regime is included. Although measurement results are available at mmWave frequencies above this range, they are intentionally excluded here because mmWave propagation is affected by substantially different blockage, scattering, atmospheric loss, and deployment characteristics. Including such bands would require a separate modeling treatment and would move the focus away from the upper-mid-band transition considered in this work.

Existing 3GPP-style models are typically parameterized independently at discrete carrier frequencies or derived by combining measurement data from separate campaigns. Such merging is unavoidable in practice, since no single organization can realistically conduct harmonized large-scale measurements over all relevant bands and scenarios. However, independent per-frequency tuning or sparsely supported cross-band merging can still lead to inconsistencies in large-scale parameters (LSPs), especially near the $7.125$~GHz where the measurements weakly represent the Upper FR1/FR3 transition region \cite{Poddar_3GPP-Rel19, Poddar_Yoshimura,[Miao]Empirical_Studies, Nokia_WP}.

Existing studies have reported useful large-scale parameter (LSP) data in Upper FR1, FR3, and low-mmWave bands, but the evidence in this transition region remains relatively sparse and fragmented across scenarios and research groups \cite{Spectrum_Sandbox}. This scarcity is not limited to outdoor environments; measurements in the $4$--$8$~GHz range are also limited for indoor industrial scenarios, although recent work has begun to address this. In particular, Ying et al.~\cite{Ying_ICC2025} reported upper-mid-band channel measurements and characterization at $6.75$~GHz FR1(C) and $16.95$~GHz FR3 in an indoor factory scenario, and angular-spread behavior reported in \cite{Yubei_Spectrum_Sandbox, NYU_AS_Indoor}, highlighting the broader need for harmonized cross-band channel knowledge beyond conventional sub-$6$~GHz deployments.
Among the LSPs commonly used in system-level and stochastic channel modeling, root-mean-square delay spread (DS), azimuth spread of arrival (ASA), azimuth spread of departure (ASD), shadow fading (SF), and Rician $K$-factor are especially relevant for describing delay dispersion, angular dispersion, and large-scale power variation. Although zenith spreads of arrival and departure (ZSA and ZSD) are important for full 3D beamforming in modern MIMO systems, the present study focuses on the azimuth-domain parameters (ASA and ASD) due to the limitations in the elevation-steering capability of the employed antenna arrays. Large-scale measurements in urban macrocell (UMa) and urban microcell (UMi) scenarios have shown that DS and angular spreads generally decrease with increasing frequency, reflecting reduced multipath richness and stronger propagation directivity at higher bands \cite{NYU_FR1_FR3}. These effects are relevant to realistic channel characterization and to the assumptions used in mobility, beam-management, and multiband simulation studies \cite{ITU-R_P.1411}. 

Nevertheless, most available models and datasets still treat FR1 and FR3 separately, or use coarse frequency binning with independently tuned parameters across bands. This practice partly reflects the practical difficulty of conducting large-scale, harmonized, multi-band measurement campaigns within a single organization or university; consequently, measurements are often collected by different research groups using different hardware, bandwidths, environments, and processing procedures. Such fragmentation can lead to non-smooth transitions across the $7.125$~GHz boundary, particularly when the $4$--$8$~GHz region is weakly represented. 

At the same time, single-band campaigns inherently reflect site-specific propagation conditions~\cite{Samsung_Research}. 
In contrast, 3GPP scenario models are generic stochastic parameterizations developed by pooling measurement and ray-tracing datasets \cite{Aalborg_Sub-6RT, 6G_material_RT}, contributed by multiple organizations, sites, cities, scenarios, and frequency bands. This aggregation is intended to reduce the site-specific bias and inter-campaign fragmentation inherent in individual measurement campaigns. For example, the recent Release~19 dataset and curve-fitting study in~\cite{Poddar_2026} documents the measurement campaigns, datasets, and fitting procedures used for TR~38.901~\cite{TR138-901} channel-model validation and refinement over the $7$--$24$~GHz range. Therefore, any difference between the present measurements and default 3GPP values should be interpreted as a mismatch between a generic scenario-level reference model and the specific propagation characteristics of the measured environment, rather than as evidence that the standard model is generally inaccurate.  Because the Release~19 data source in~\cite{Poddar_2026} mainly targets the $7$--$24$~GHz range, the present $4.85$~GHz measurements provide complementary Upper FR1 outdoor urban evidence below that range, in a deployment-relevant band near the lower side of the Upper FR1/FR3 transition.

To address these limitations, we present a $4.85$~GHz measurement-anchored urban channel study together with a literature-referenced cross-band trend analysis for key LSPs. The carrier frequency of $4.85$~GHz was primarily determined by the available in-house Sub-6 GHz MIMO channel sounder, which was designed and calibrated for this band. Nevertheless, this frequency is technically relevant because it lies in the upper part of FR1, close to the FR3 transition region, and provides evidence from outdoor urban measurements in a frequency range where harmonized double-directional datasets remain limited. 

Specifically, we conduct a precisely calibrated double-directional measurement campaign at $4.85$~GHz in three urban routes in Yokohama, Japan, covering UMa and UMi conditions. From these data, we extract path loss (PL), DS, ASA, ASD, $K$-factor, and route-dependent spatial-consistency statistics.
\begin{figure}[t!]
    \centering
    \includegraphics[width=\linewidth]{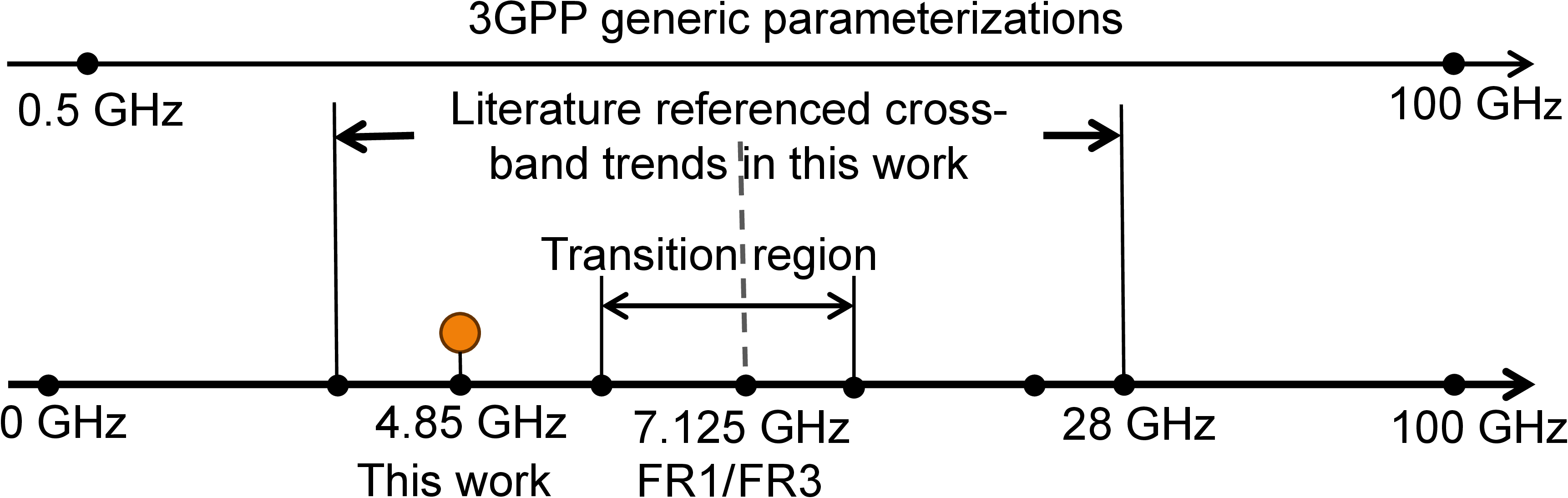}
 \caption{Study scope: direct measurements at $4.85$~GHz and literature-referenced cross-band context from approximately $4$ to $28$~GHz, focused on the Upper FR1/FR3 transition and excluding FR2/mmWave channel behavior}
 \label{FrequencyTimeline}
\end{figure}
To place the $4.85$~GHz results in a broader context, we combine route-wise mean LSP values from the present campaign with scenario-matched literature anchors up to $28$~GHz, as illustrated in Fig.~\ref{FrequencyTimeline}, and derive log-log trends for DS, ASA, and ASD. These trends are compared with the 3GPP UMa/UMi reference parameterizations \cite{TR138-901} over the same anchor-supported frequency interval. In this way, the direct experimental results of this paper are obtained at $4.85$~GHz, while the broader Upper FR1/FR3 perspective is addressed through this literature-referenced cross-band analysis. However, because this part relies on a single in-house measurement band together with a limited and heterogeneous set of literature anchors, it is presented only as indicative and measurement-informed, not as a definitive multi-band model or as a basis for direct recommendations to 3GPP or ITU. A genuinely smooth and fully validated cross-band parameterization would require harmonized measurements across multiple bands, environments, and regions \cite{Zhang6G_Modeling, Unified6GModel, Zhang6G_Tutorial}.

Our framework also incorporates a distance-based spatial-consistency analysis that extracts route-specific decorrelation distances for shadow fading (SF), obtained from the residual between the measured PL and the fitted large-scale PL trend, as well as for key LSPs, providing empirical insight into how large-scale channel characteristics evolve along measured trajectories. These results are relevant to spatially consistent stochastic simulation and mobility-related evaluations such as handover and beam tracking. However, the reported decorrelation distances should be interpreted as route-dependent empirical indicators of local spatial persistence rather than universal scenario constants. Thus, the use of $4.85$~GHz reflects both the practical availability of a calibrated measurement platform and the relevance of this Upper FR1 band for assessing continuity toward lower-FR3 channel behavior.

The main contribution of this paper is therefore a measurement-based, route-resolved characterization of Upper FR1 urban propagation at $4.85$~GHz, rather than a new universal channel model. Specifically, the paper provides:
\begin{itemize}
\item A double-directional measurement-based characterization of urban channels at $4.85$~GHz in three UMa/UMi routes, providing PL and key LSP statistics in a sparsely studied Upper FR1 band;
\item A route-wise CI/FI PL fitting and comparison with ITU-R and 3GPP references, showing both broad consistency with standardized models and geometry-dependent deviations in specific LoS/NLoS regions;
\item Empirical SF and LSP spatial-consistency statistics that describe how large-scale channel behavior evolves along measured urban trajectories; and
\item A bounded literature-referenced comparison over approximately $4$--$28$~GHz to place the $4.85$~GHz results in Upper FR1/FR3 context without claiming a definitive multi-band parameterization.
\end{itemize}

The remainder of this paper is organized as follows: Section~II describes the measurement campaign and environments. Section~III presents the channel extraction and processing methodology. Section~IV reports the measured channel characteristics and route-based spatial-consistency analysis. Section~V provides the literature-referenced cross-band trend analysis. Finally, Section~VI concludes the paper.
\begin{figure}[t!]
    \centering
        \vspace{1mm}
    \includegraphics[width=0.7\linewidth]{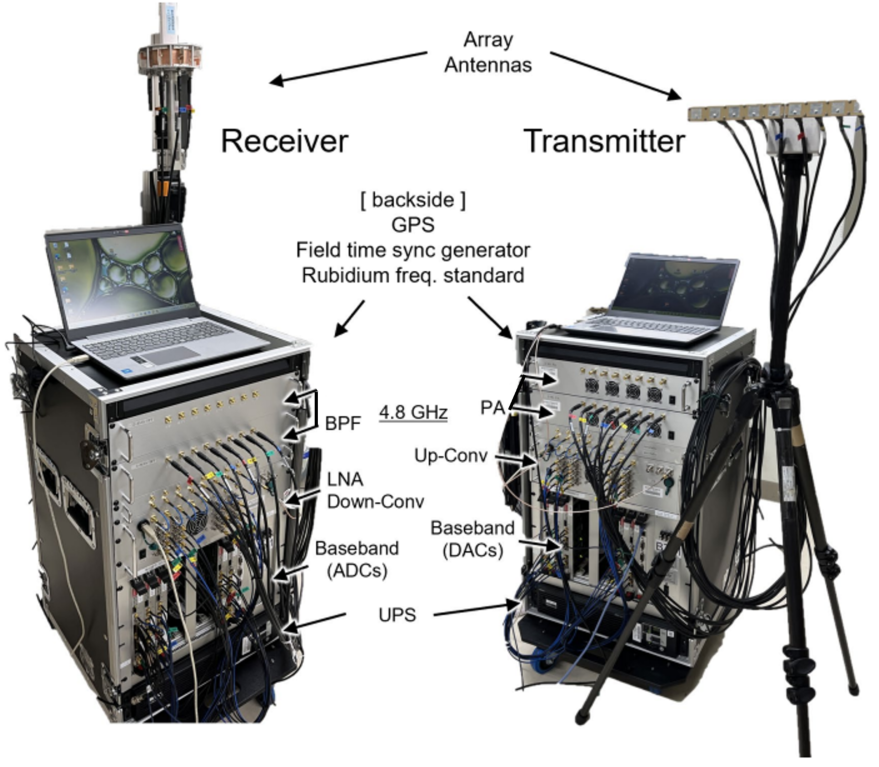}
 \caption{Measurement system.}
 \label{ChannelSounder}
\end{figure}
\begin{figure*}[t]
\centering
\subfigure[{\tt Area1} (UMa).\label{YurakuArea}]{\includegraphics[width=0.32\linewidth]{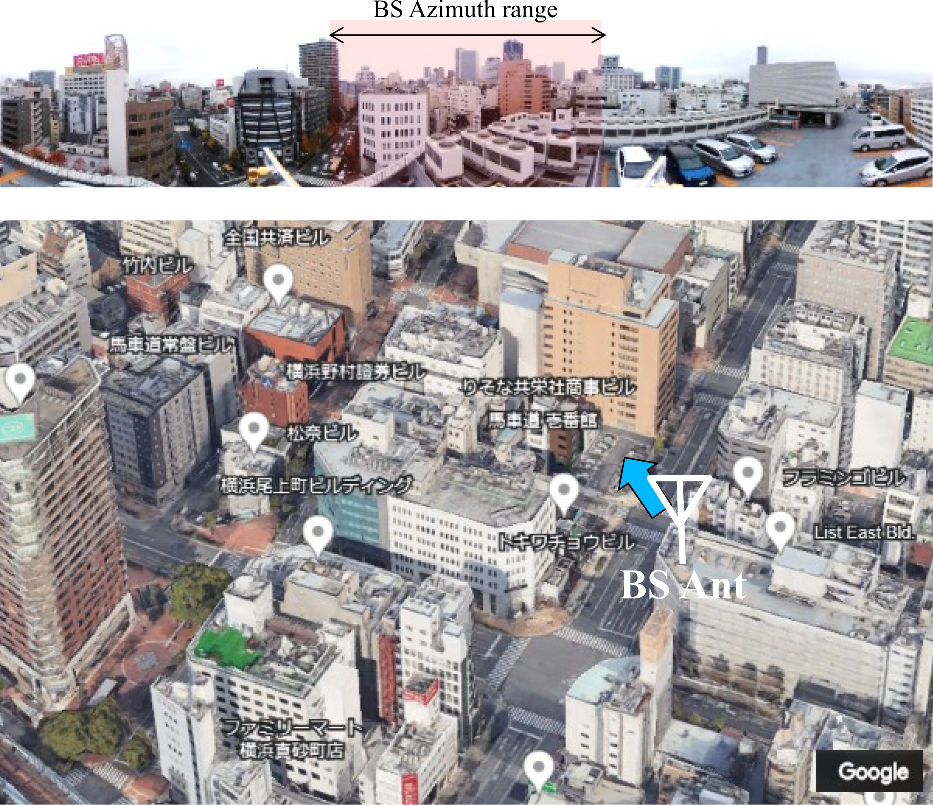}} 
\subfigure[{\tt Area2} (UMa).\label{NTTArea}]{\includegraphics[width=0.32\linewidth]{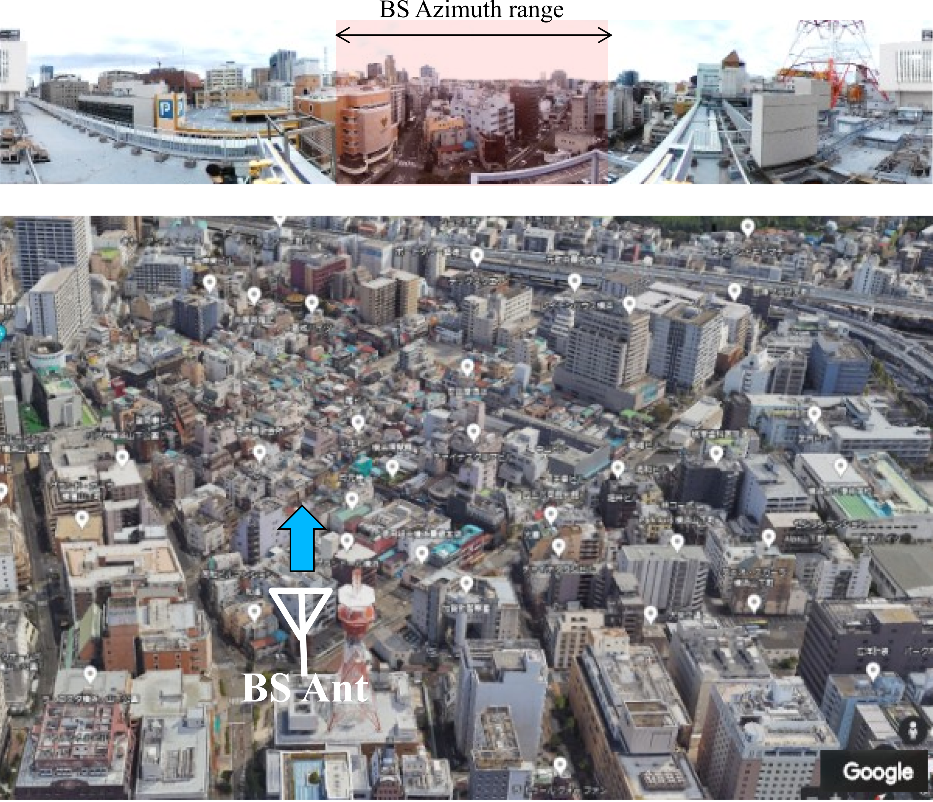}} 
\subfigure[{\tt Area3} (UMi).\label{WPArea}]{\includegraphics[width=0.32\linewidth]{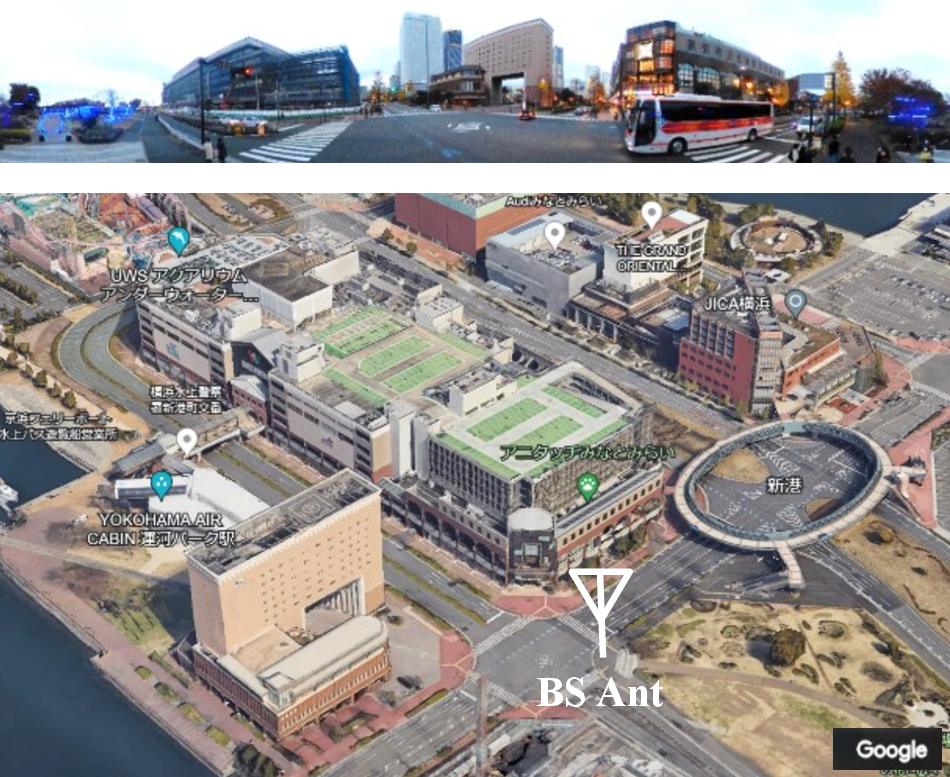}} 
\caption{Three urban cellular scenarios. \label{three_scenarios}}
\end{figure*}
\begin{table*}[t]
\centering
\caption{Measurement setups.}
\label{table_scenarios}

\renewcommand{\arraystretch}{1.15}
\setlength{\tabcolsep}{4pt}

\begin{tabular}{>{\centering\arraybackslash}m{2.2cm}|m{6.2cm}|>{\centering\arraybackslash}m{2.5cm}>{\centering\arraybackslash}m{2.2cm}|>{\centering\arraybackslash}m{2.8cm}}
\hline
Scenarios & Area & \multicolumn{2}{c|}{BS Antennas} & MS Antennas \\ \hline
\multirow{12}{*}{\begin{tabular}[c]{@{}c@{}}Urban Macro\\ (UMa)\end{tabular}}
& \begin{tabular}[l]{@{}l@{}}
- Kannai Area ({\tt Area1}), \\
- BS Location: Yokohama Daiichi Yuraku Bldg, \\
- Meas. Points: 137 (LoS) and 2,863 (NLoS), \\
- Valid Points: 69 (LoS) and 846 (NLoS)
\end{tabular}
& \begin{tabular}[c]{@{}c@{}}8-elem ULA\\ Height: $33$ m\\ Direction: N\end{tabular}
& \begin{minipage}{20mm}
\centering
\includegraphics[width=\linewidth]{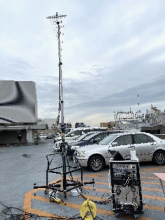}
\end{minipage}
& \multirow{3}{*}{\begin{tabular}[c]{@{}c@{}}
8-elem UCA\\
Height: $2.7$ m\\
(on vehicle roof) \\
\begin{minipage}{20mm}
\centering
\scalebox{1.3}{\includegraphics[width=\linewidth]{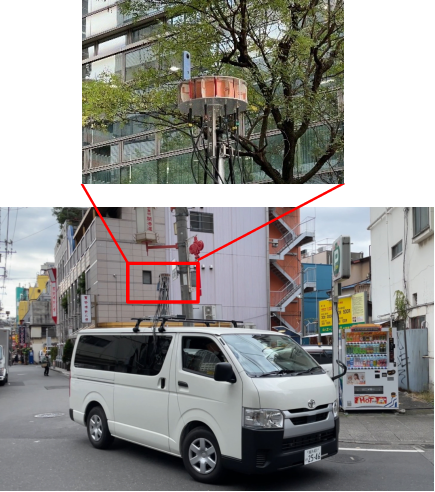}}
\end{minipage}
\end{tabular}}
\\ \cline{2-4}

& \begin{tabular}[l]{@{}l@{}}
- Chinatown ({\tt Area2}), \\
- BS Location: NTTCom Yokohama Yamashita Bldg, \\
- Meas. Points: 243 (LoS) and 2,757 (NLoS), \\
- Valid Points: 182 (LoS) and 452 (NLoS)
\end{tabular}
& \begin{tabular}[c]{@{}c@{}}8-elem ULA\\ Height: $34$ m\\ Direction: SE\end{tabular}
& \begin{minipage}{20mm}
\centering
\includegraphics[width=\linewidth]{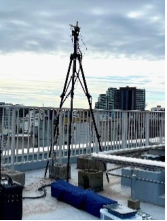}
\end{minipage}
& \\ \cline{1-4}

\multirow{4}{*}{\begin{tabular}[c]{@{}c@{}}Urban Micro\\ (UMi)\end{tabular}}
& \begin{tabular}[l]{@{}l@{}}
- Streets ({\tt Area3}), \\
- BS Location: Yokohama World Porters, \\
- Meas. Points: 1,419 (LoS) and 1581 (NLoS), \\
- Valid Points: 827 (LoS) and 908 (NLoS)
\end{tabular}
& \begin{tabular}[c]{@{}c@{}}8-elem UCA\\ Height: $3.0$ m\end{tabular}
& \begin{minipage}{20mm}
\centering
\includegraphics[width=\linewidth]{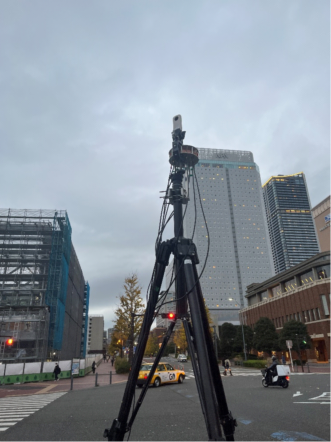}
\end{minipage}
& \\ \hline
\end{tabular}
\end{table*}

\section{Measurement Campaign and Setup}

\subsection{Measurement System}
Channel measurements were carried out using an in-house-developed Sub-$6$~GHz ($4.85$~GHz band) multiple-input-multiple-output (MIMO) channel sounder operating at a center frequency of $4.85001$~GHz, shown in Fig.~\ref{ChannelSounder}. The sounder includes custom RF frequency-conversion circuits and antenna arrays optimized for this band \cite{MIMOCHSounder}.  An unmodulated Newman-phase multitone waveform with $N=510$ tones over a $99.9$~MHz bandwidth was used for channel sounding, yielding a nominal delay resolution of approximately $10$~ns, a tone spacing of $195$~kHz, and a maximum delay span of $5.12~\mu$s.

Before the measurement campaign, the sounder underwent an end-to-end validation and calibration procedure including software verification, transmitter/receiver (Tx/Rx) IQ imbalance checks, back-to-back (B2B) measurements, single-shot verification, time-grid verification, GPS-linked timing checks, and a full calibration sequence \cite{IEICE_Kim}. Following end-to-end calibration, the sounder achieved an effective dynamic range of approximately $50$~dB. Together with the measured Rx noise floor, this was sufficient for stable wideband channel transfer function (CTF) acquisition over the urban link distances covered in this study.

The Tx and Rx were configured in an $8 \times 8$ full-MIMO architecture to enable double-directional wideband channel characterization. The Tx, representing the base station (BS), employed an 8-element uniformly spaced linear array (ULA) with vertically polarized elements. The Rx, representing the mobile station (MS), employed an 8-element uniformly spaced circular array (UCA) to provide full azimuthal coverage. 
The individual array elements had broad element patterns rather than omnidirectional patterns: the ULA elements exhibited a half-power beamwidth (HPBW) of approximately \ang{90} with a gain of about $4$~dBi, whereas the UCA elements had an HPBW of approximately \ang{74} and a gain of about $6.5$~dBi. The narrower effective beams used later in the directional post-processing are synthesized by array-domain spatial filtering and should therefore be distinguished from the broader element patterns described here. The full-MIMO architecture used one RF chain per antenna element, enabling parallel transmission and reception. Signal multiplexing was achieved through a hybrid scheme combining frequency-division multiplexing (FDM) and space-time division multiplexing (STDM) with orthogonal coding \cite{MIMOCHSounder}, which minimizes inter-element interference and supports reliable multipath component (MPC) extraction.
\begin{figure*}[t]
\centering
\subfigure[{\tt Area1} (UMa).\label{Yuraku_Route}]{\includegraphics[width=0.32\linewidth]{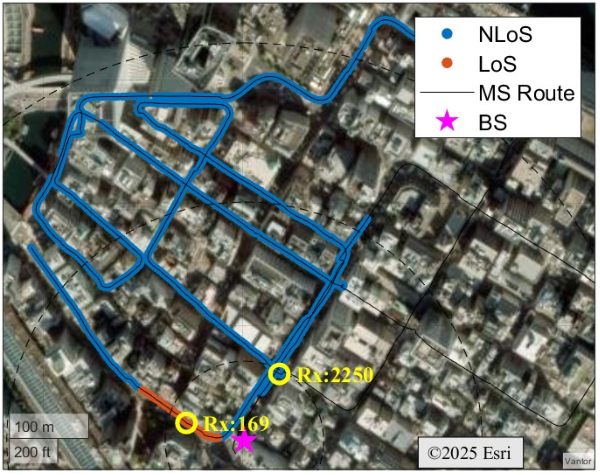}} 
\subfigure[{\tt Area2} (UMa).\label{NTT_Route}]{\includegraphics[width=0.32\linewidth]{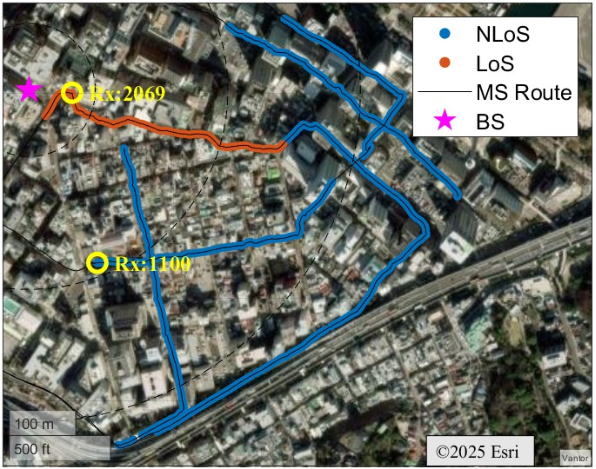}} 
\subfigure[{\tt Area3} (UMi).\label{WP_Route}]{\includegraphics[width=0.32\linewidth]{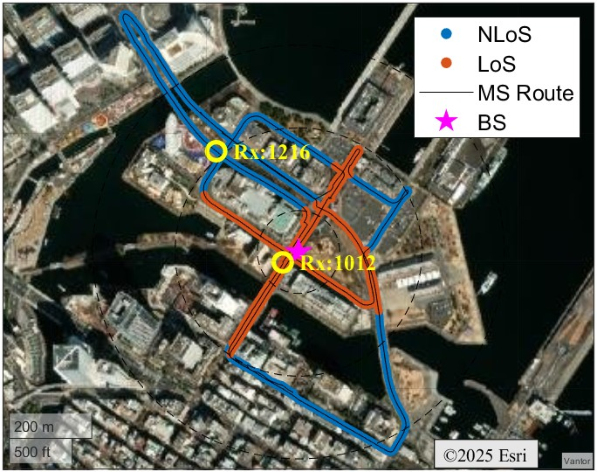}} 
\caption{Measurement routes and representative Rx positions used for the example PDPs, showing BS location and LoS/NLoS route segmentation.
\label{MeasRoute}}
\end{figure*}

\subsection{Test Environments and Route Scenarios}
Extensive channel measurements were conducted in three representative urban areas of Yokohama City, Japan, representing the UMa and UMi deployment types considered in 3GPP-style studies \cite{3GPP-TR138_921}. The three measurement sets are denoted as {\tt Area1}, {\tt Area2}, and {\tt Area3} in Fig.~\ref{three_scenarios}.

{\tt Area1} and {\tt Area2} correspond to UMa-like deployments, where the BS antenna is located above the surrounding rooftop level. In {\tt Area1} (Fig.~\ref{YurakuArea}), an 8-element ULA was mounted at approximately $33$~m height near a railway-station district, whereas in {\tt Area2} (Fig.~\ref{NTTArea}), an 8-element ULA was installed at approximately $34$~m height in the densely built Chinatown area. The BS boresights were oriented approximately northward in {\tt Area1} and southeastward in {\tt Area2}. In both cases, the BS array provided a fixed azimuth sector of about \ang{-50} to \ang{+50}, as illustrated in Fig.~\ref{three_scenarios}, and the propagation is mainly influenced by rooftop diffraction, corner interaction, and urban-canyon reflections typical of macrocell deployments.
{\tt Area3} corresponds to a UMi-like deployment near a shopping complex district, where the BS used an 8-element UCA mounted at approximately $3$~m height, as shown in Fig.~\ref{WPArea}. In this case, the propagation is governed primarily by street-canyon and around-building mechanisms, with much less sensitivity to rooftop-level effects. In all campaigns, the MS employed a UCA mounted on the roof of the measurement vehicle at approximately $2.7$~m height. The main BS/MS antenna configurations are summarized in Table~\ref{table_scenarios}.
CTF snapshots were collected while the measurement vehicle moved along typical urban streets at speeds below $20$~km/h under realistic traffic conditions. Snapshots were recorded every $0.5$~s, corresponding to a nominal spatial increment of approximately $2.75$~m at the maximum speed of $20$~km/h (about $5.5$~m/s), although the actual spacing varied slightly with traffic conditions. Approximately $3000$ raw snapshots were obtained in each area. The routes shown in Fig.~\ref{MeasRoute} include mixed building facades, vehicular traffic, pedestrian activity, and seasonal roadside vegetation, thereby providing a realistic basis for analyzing route-dependent channel characteristics in dense urban environments.

\subsection{Environment Characterization}
For the macrocell measurements in {\tt Area1} and {\tt Area2}, the BS azimuth coverage was limited to approximately \ang{-50} to \ang{+50}, as shown in Fig.~\ref{three_scenarios}. Because the macrocell BS sectors in {\tt Area1} and {\tt Area2} were limited to an azimuth range of approximately \ang{-50} to \ang{+50}, some MS positions fell outside the intended sector coverage during the drive tests. Measurements from such positions were excluded from further analysis. Accordingly, the retained dataset should be interpreted as a sector-specific measurement set rather than a full omnidirectional site-level characterization.
The remaining valid snapshots were further classified into LoS and NLoS segments offline by cross-referencing the time-synchronized GPS coordinates of the measurement vehicle with a 3D digital map to determine whether the direct BS--MS path was optically blocked for each snapshot. This geometry-based classification was additionally verified using vehicle-mounted panoramic camera footage. The resulting LoS/NLoS intervals are indicated in Fig.~\ref{MeasRoute}.
To reduce oversampling when the vehicle stopped or moved only marginally, consecutive snapshots were screened according to spatial separation. If the distance between adjacent snapshots was less than $1.0$~m, the later snapshots were treated as duplicates, and only the first was retained. This procedure suppresses spatial oversampling caused by traffic lights and temporary stops while preserving the large-scale route evolution. The remaining valid samples comprise $69/846$, $182/452$, and $827/908$ LoS/NLoS positions in {\tt Area1}, {\tt Area2}, and {\tt Area3}, respectively.
\begin{figure}[t]
\centering
\includegraphics[width=0.999\linewidth]{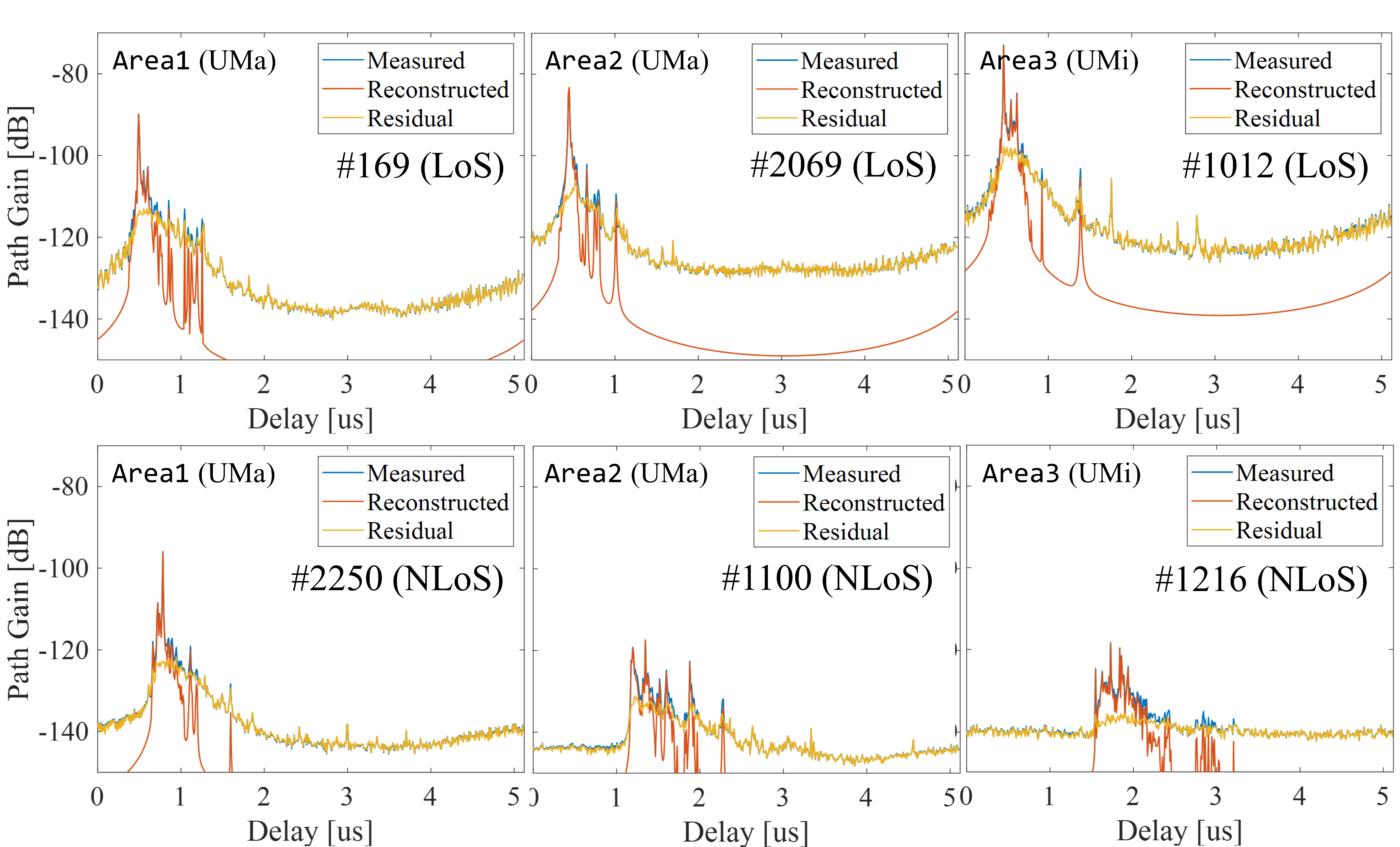}
\caption{Power delay profiles (PDPs) for example representative LoS/NLoS positions computed from the measured, MPC-reconstructed, and residual MIMO channel matrices using \eqref{eq:PDP}. 
\label{fig:PDPs}}
\end{figure}

\section{Channel Extraction}

\subsection{MIMO Channel Matrix and Multidimensional Power Spectrum}
The measured wideband channel at each snapshot is represented by the MIMO channel transfer function (CTF) matrix $\Vect{H}_{\mathrm{MIMO}}(f)\in\mathbb{C}^{8\times 8}$. Because the measured response is shaped by the Tx/Rx array responses, directional post-processing is used to obtain an array-compensated double-directional representation in the delay-azimuth domain. This operation reduces, but does not completely remove, the influence of the element and synthesized beam patterns; in particular, residual sidelobe leakage cannot be fully avoided. 

Let the Tx and Rx azimuth pointing angles be defined as
\begin{eqnarray}
\check{\varphi}_\T \in \{\, n_{\varphi_\T}\Delta_{\varphi_\T}
    \mid n_{\varphi_\T} = 0, \ldots, N_{\varphi_\T} - 1 \,\},\\
\check{\varphi}_\R \in \{\, n_{\varphi_\R}\Delta_{\varphi_\R}
    \mid n_{\varphi_\R} = 0, \ldots, N_{\varphi_\R} - 1 \,\},
\end{eqnarray}
where $\Delta_{\varphi_\T}$ and $\Delta_{\varphi_\R}$ denote the angular resolutions at the Tx and Rx, respectively.
Using the array response vectors $\Vect{a}_\T(\check{\varphi}_\T) \in \mathbb{C}^{8\times 1}$ and $\Vect{a}_\R(\check{\varphi}_\R) \in \mathbb{C}^{8\times 1}$, the double-directional CTF is obtained as
\begin{multline}
G(f,\check{\varphi}_\T,\check{\varphi}_\R)  =  \\ C_{\R}^{-1}(\check{\varphi}_\R)\,  \Vect{a}_\R^{H}(\check{\varphi}_\R)\, \Vect{H}_{\mathrm{MIMO}}(f)\,  \Vect{a}_\T(\check{\varphi}_{\T})\,  C_\T^{-1}(\check{\varphi}_{\T}),
\label{eq:BF}
\end{multline}
where the compensation coefficients for the antenna gains are defined as $C_\T(\check{\varphi}_{\T}) = \Vect{a}_{\T}^H(\check{\varphi}_\T)\Vect{a}_{\T}(\check{\varphi}_\T)$, and $C_\R(\check{\varphi}_\R) = \Vect{a}_{\R}^H(\check{\varphi}_\R)\Vect{a}_{\R}(\check{\varphi}_\R)$.
The corresponding double-directional impulse response, denoted by $g(\tau,\check{\varphi}_\T,\check{\varphi}_\R)$, is obtained by applying the inverse Fourier transform to $G(f,\check{\varphi}_\T,\check{\varphi}_\R)$ with respect to frequency \cite{IEICE_Kim}. The channel can be regarded as effectively static within a single snapshot: at the maximum vehicle speed of $20$~km/h, the Doppler shift at $4.85$~GHz is about $90$~Hz, corresponding to a coherence time of roughly $11$~ms, which is much longer than the $5.12~\mu$s duration of the sounding waveform.
The angular-delay power spectrum is then expressed as
\begin{eqnarray}
P(\tau,\check{\varphi}_\T,\check{\varphi}_\R) = \left|g(\tau,\check{\varphi}_\T,\check{\varphi}_\R)\right|^2.
\label{eq:P}
\end{eqnarray}
The power delay profile (PDP) is computed from \eqref{eq:P} as
\begin{eqnarray}
P_{h}(\tau) =  \frac{\Delta_{\varphi_\T}\Delta_{\varphi_\R}}{B_{\varphi_\T} B_{\varphi_\R}} \sum_{\check{\varphi}_\T} \sum_{\check{\varphi}_\R} P(\tau,\check{\varphi}_\T,\check{\varphi}_\R),
\label{eq:PDP}
\end{eqnarray}
where $B_{\varphi_\T}$ and $B_{\varphi_\R}$ denote the effective HPBWs of the synthesized Tx and Rx beamformers, respectively. In this work, the angular sampling steps were set to $\Delta_{\varphi_\T} = \Delta_{\varphi_\R} = 6^\circ$. The resulting beamwidths were approximately $12^\circ$ for the 8-element ULA and $24^\circ$ for the 8-element UCA.

\subsection{Multipath Components (MPC) Extraction and Clustering}

The measured MIMO channel impulse response (CIR) contains the propagation channel convolved with the effective hardware and array responses, including the Tx/Rx radiation patterns. It can be written as     
\begin{multline}
  \Vect{h}_{\mathrm{MIMO}}(\tau)  =  \iiint\limits_{\kappa, \varphi_\T, \varphi_\R} \Vect{a}_\R(\varphi_\R)\, h_c(\tau, \varphi_\T, \varphi_\R)\,  \Vect{a}_\T^{T}(\varphi_\T)\cdot \\ a_u(\tau - \kappa)\, d\kappa\, d\varphi_\T\, d\varphi_\R =  \mathcal{F}^{-1} \Vect{H}_{\mathrm{MIMO}}(f),
\label{eq:Measured_h1} 
\end{multline}
where $\tau$ denotes the delay, $\varphi_{\T}$ the angle of departure, and $\varphi_{\R}$ the angle of arrival, $a_u(\cdot)$ the autocorrelation of the sounding waveform \cite{DelayTracking}, and $h_c(\kappa,\varphi_\T,\varphi_\R)$ the underlying propagation CIR. $\Vect{a}_\T(\varphi_{\T})$ and $\Vect{a}_\R(\varphi_{\R})$ are the Tx/Rx antenna array radiation patterns, respectively.
Assuming a discrete superposition of $L$ plane-wave components \cite{Kim_Springer}, the propagation CIR is modeled as
\begin{multline}
   h_c(\tau, \varphi_\T, \varphi_\R) =  \\ \sum_{l=1}^{L} \gamma_l \,  \delta(\tau - \tau_l) \,  \delta(\varphi_\T - \varphi_{\T,l}) \,  \delta(\varphi_\R - \varphi_{\R,l}),
\label{eq:ParametricModel}  
\end{multline}
where $\gamma_l$, $\tau_l$, $\varphi_{\T,l}$, and $\varphi_{\R,l}$ denote the complex path weight, delay, angle of departure, and angle of arrival of the $l$th component, respectively, and $\delta(\cdot)$ denotes the Dirac delta function. Substituting \eqref{eq:ParametricModel} into \eqref{eq:Measured_h1} gives
\begin{equation}
\Vect{h}_{\mathrm{MIMO}}(\tau) = \sum_{l=1}^{L} \gamma_l a_u(\tau - \tau_l) \Vect{a}_\R(\varphi_{\R,l}) \Vect{a}_\T^T(\varphi_{\T,l}).
\label{eq:h_MIMO}
\end{equation} 

To estimate the MPC parameters from \eqref{eq:h_MIMO}, we employ the SAGE (Space-Alternating Generalized Expectation-Maximization) algorithm \cite{SAGE}, which is widely used for high-resolution delay-angle parameter estimation. In the present implementation, up to $300$ MPCs were extracted per snapshot to capture the dominant specular and quasi-specular contributions while keeping the residual unexplained energy small. Fig.~\ref{fig:PDPs} compares example PDPs obtained from the measured channel, the MPC-reconstructed channel, and the residual channel for representative positions along the routes in Fig.~\ref{MeasRoute}. After subtracting the reconstructed contribution of the estimated MPCs from the measured channel, the residual power for the representative LoS examples shown is about $5$\%, indicating that the dominant resolvable components were well captured. For the representative NLoS examples, the residual power is roughly in the $10$--$40$\% range, which is higher than in LoS. This is because the residual reflects both unmodeled diffuse multipath and non-negligible measurement noise that are not captured by the finite reconstructed MPC set. As illustrated in Fig.~\ref{fig:PDPs}, the reconstructed channel still captures the principal delay region and main reconstructed peaks in the representative NLoS cases, although the residual energy becomes more pronounced at some locations. The extracted SAGE-based MPC set should therefore be interpreted as a compact parametric representation of the dominant channel structure, with higher residual energy naturally expected in more complex NLoS conditions.
Since many stochastic channel models adopt a cluster-based representation, the extracted MPCs were subsequently grouped using the K-power-means (KPM) clustering algorithm \cite{IEICE_Kim}. In this work, clustering is primarily used as an auxiliary step for channel interpretation and consistency with cluster-based modeling practice; the primary focus remains the LSP extraction presented in Section IV.
\begin{figure*}[t]
\centering
\subfigure[{\tt Area1} (UMa).\label{fig:PL_Fit_R3}]{\includegraphics[width=0.325\linewidth]{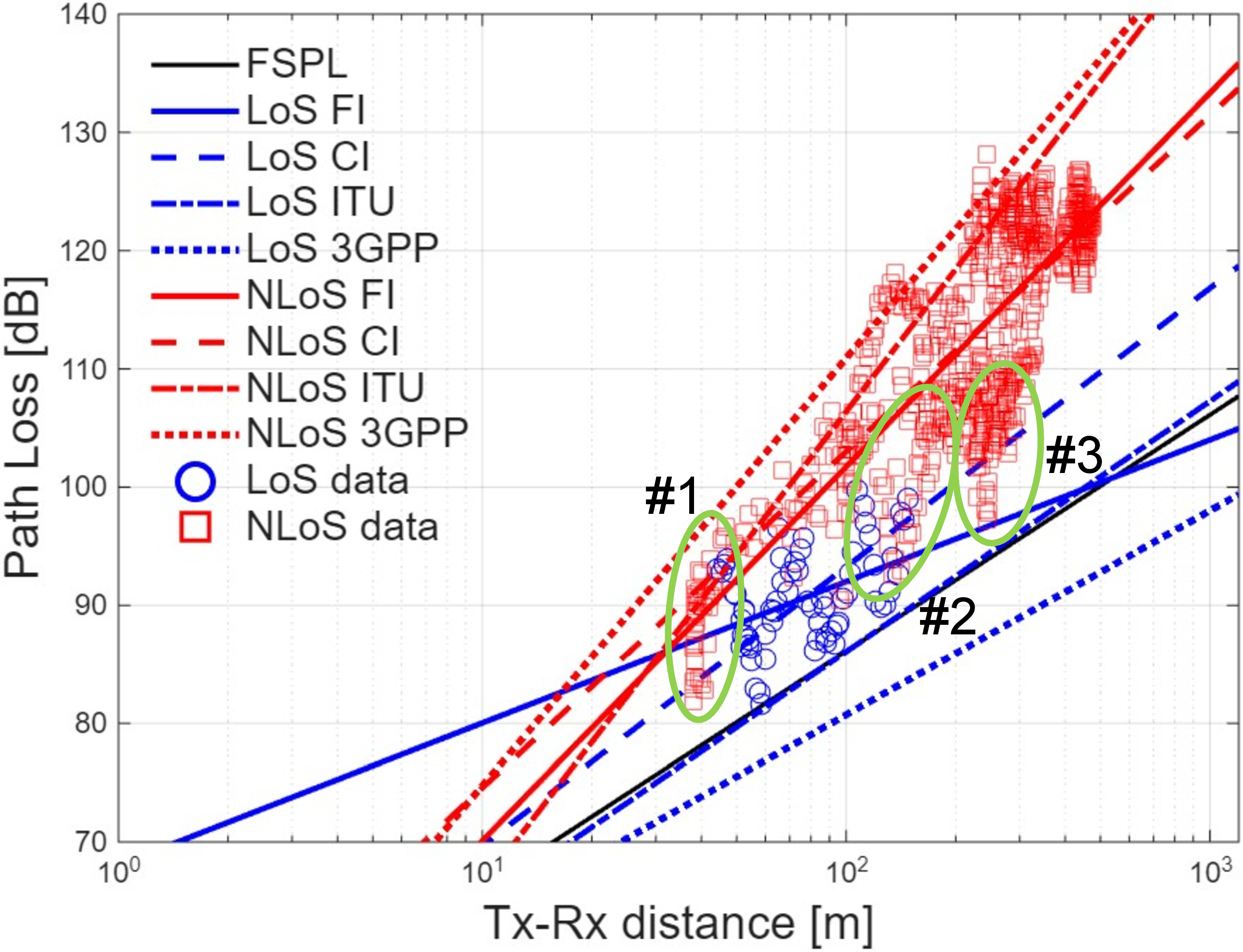}} 
\subfigure[{\tt Area2} (UMa).\label{fig:PL_Fit_R4}]{\includegraphics[width=0.325\linewidth]{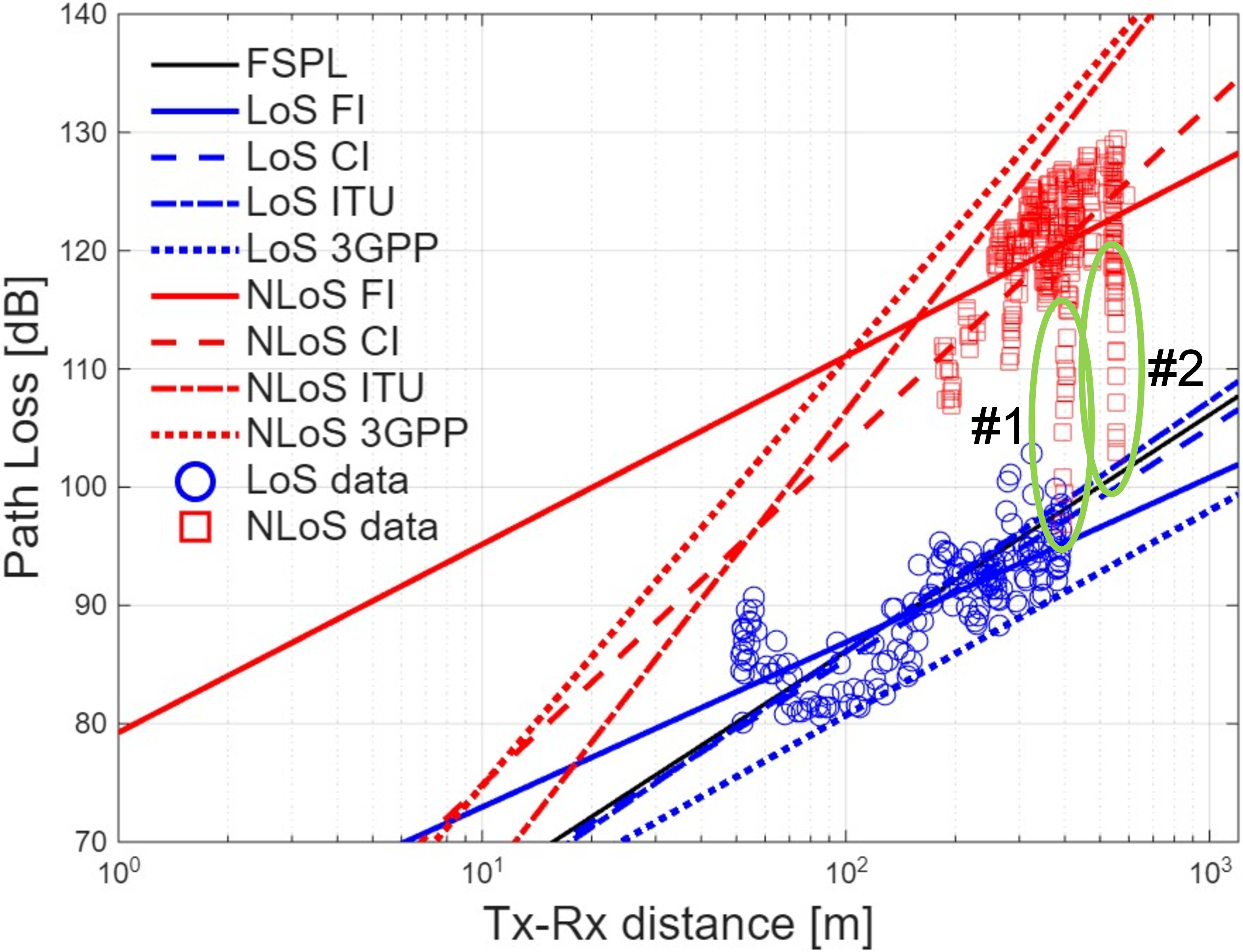}} 
\subfigure[{\tt Area3} (UMi).\label{fig:PL_Fit_R6}]{\includegraphics[width=0.325\linewidth]{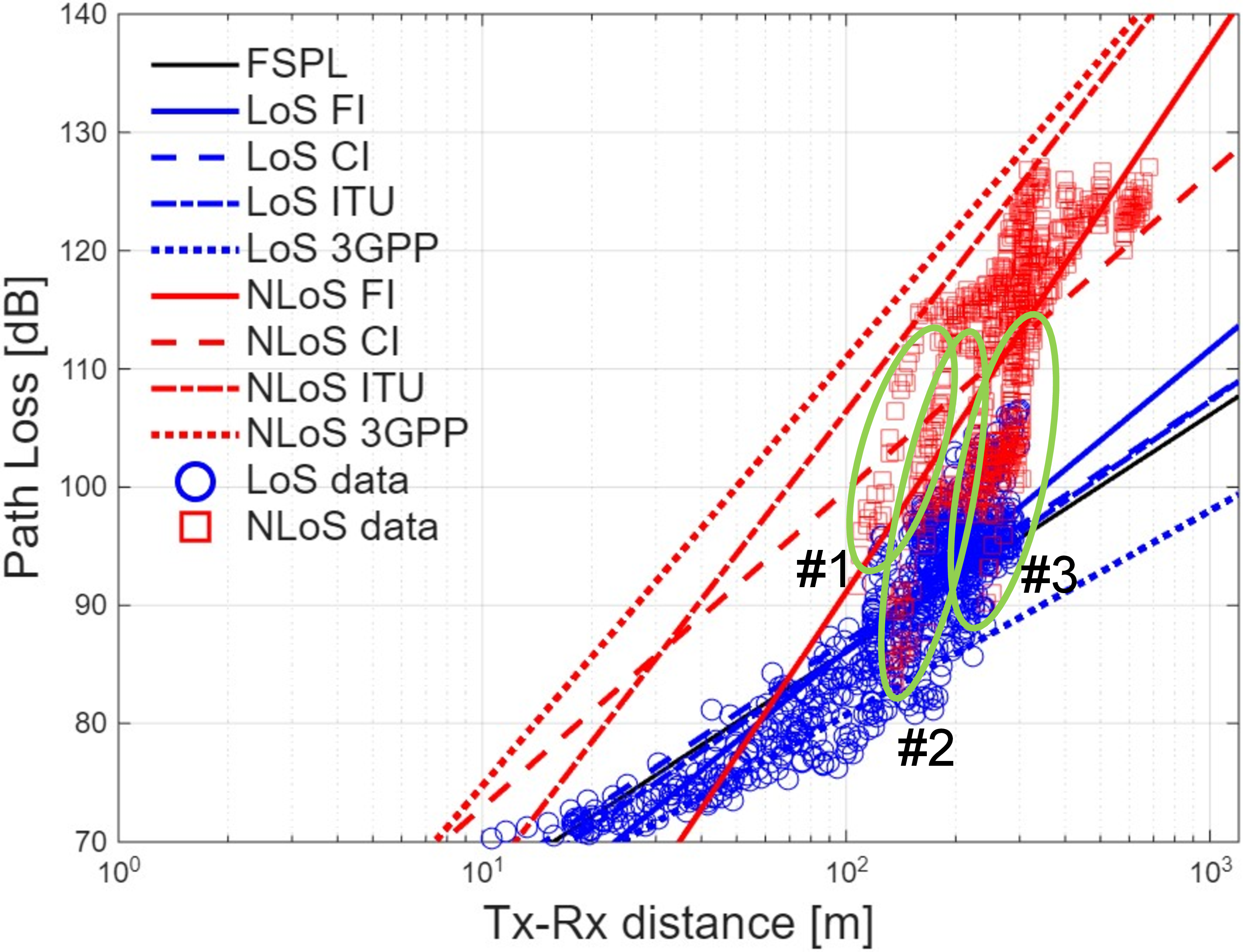}} 
\caption{PL modeling results with corresponding CI and FI regression curves plotted for LoS and NLoS, together with the FSPL, ITU-R, and 3GPP TR~38.901~\cite{TR138-901} reference curves. The ITU-R and 3GPP curves are included only as standardized scenario-level references and are not fitted to the measured data. The circled regions labeled \#1--\#3 mark representative local deviations from the dominant distance trend, associated with route-dependent geometry effects such as corners, intersections, and LoS--NLoS transition regions.} 
\label{PL_fit}
\end{figure*}
\begin{figure*}[t]
\centering
\subfigure[{\tt Area1} (UMa) Route.\label{fig:R3_PL}]{\includegraphics[width=0.30\linewidth]{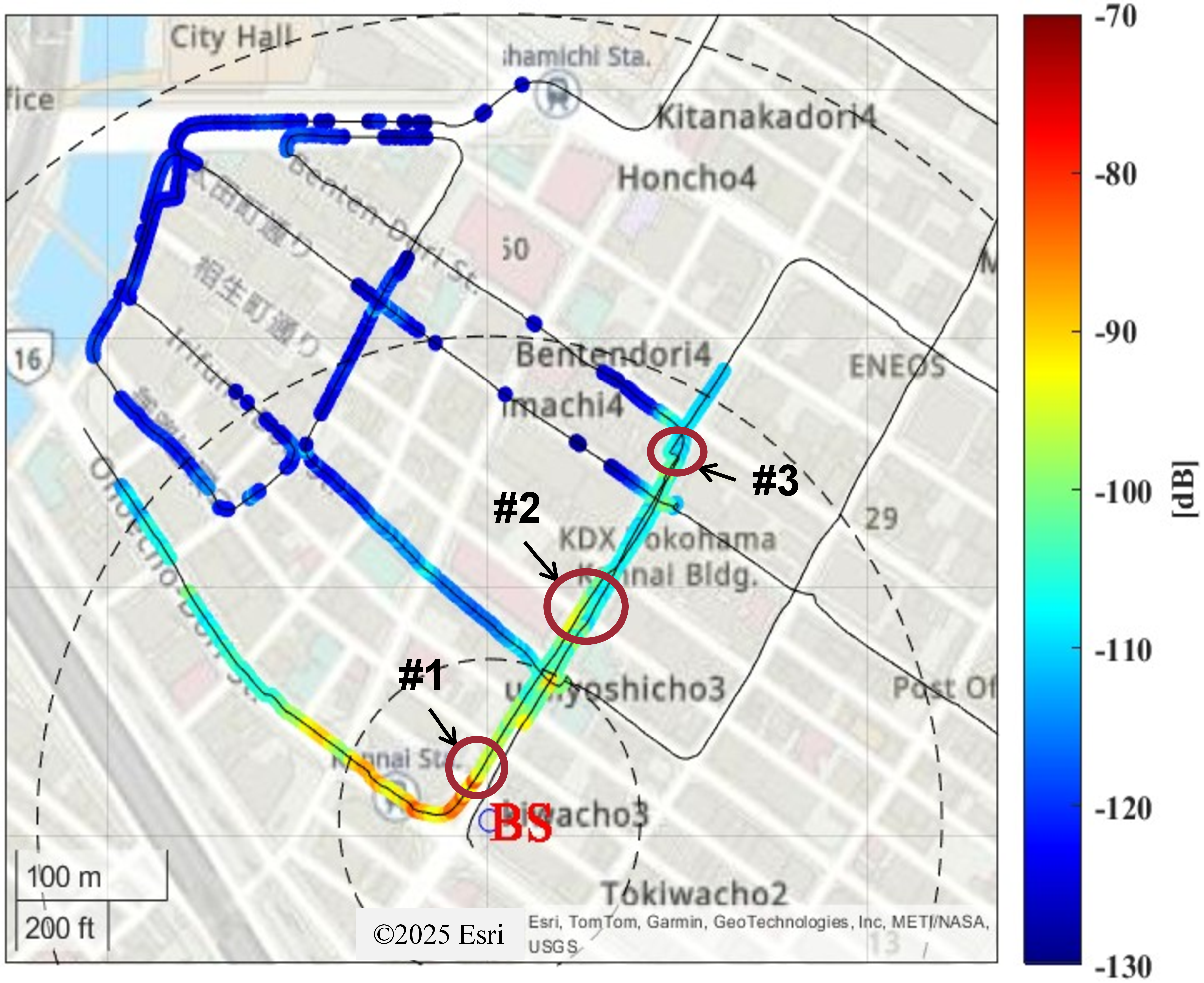}} 
\subfigure[{\tt Area2} (UMa) Route.\label{fig:R4_PL}]{\includegraphics[width=0.335\linewidth]{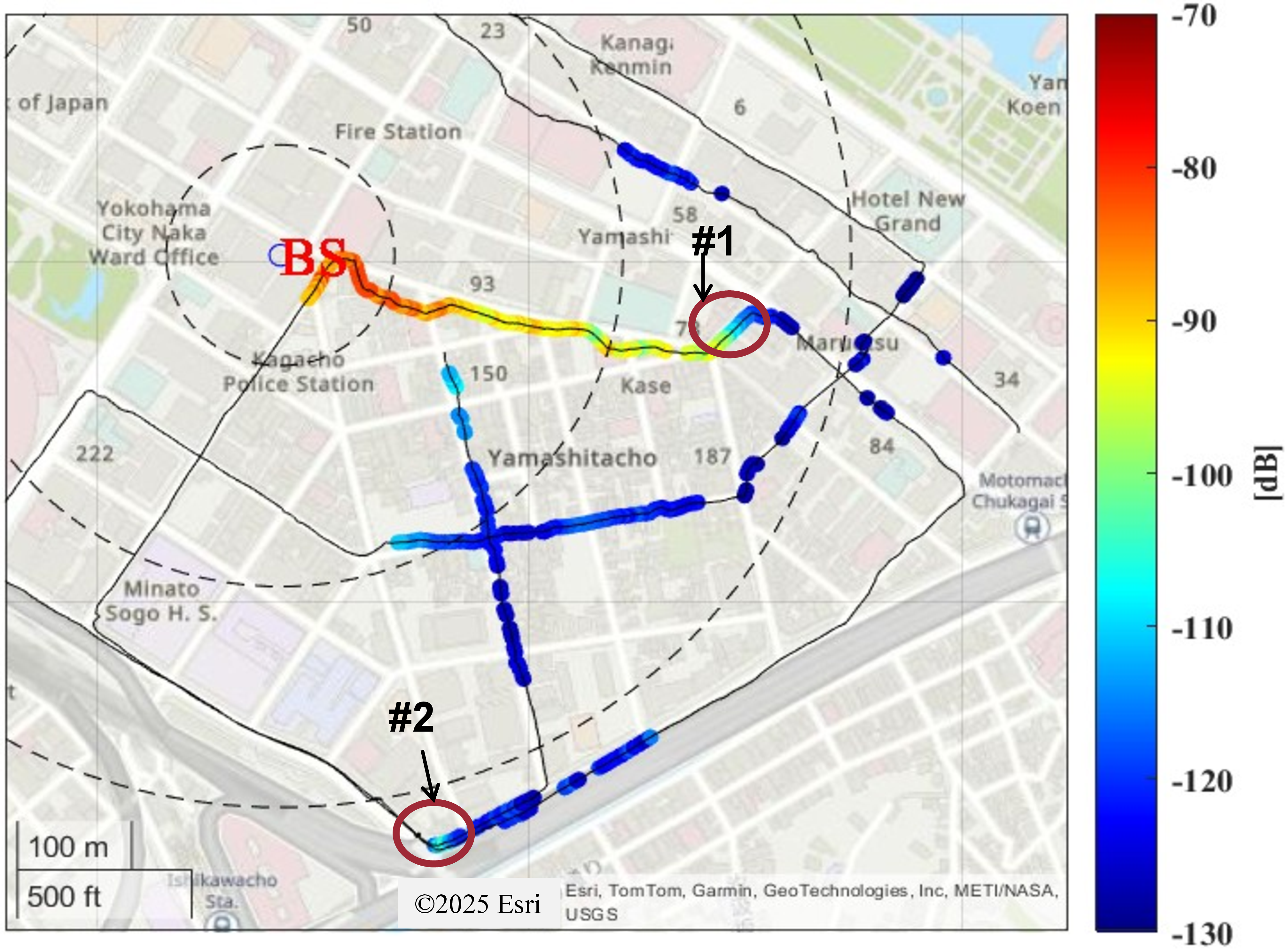}} 
\subfigure[{\tt Area3} (UMi) Route.\label{fig:R6_PL}]{\includegraphics[width=0.325\linewidth]{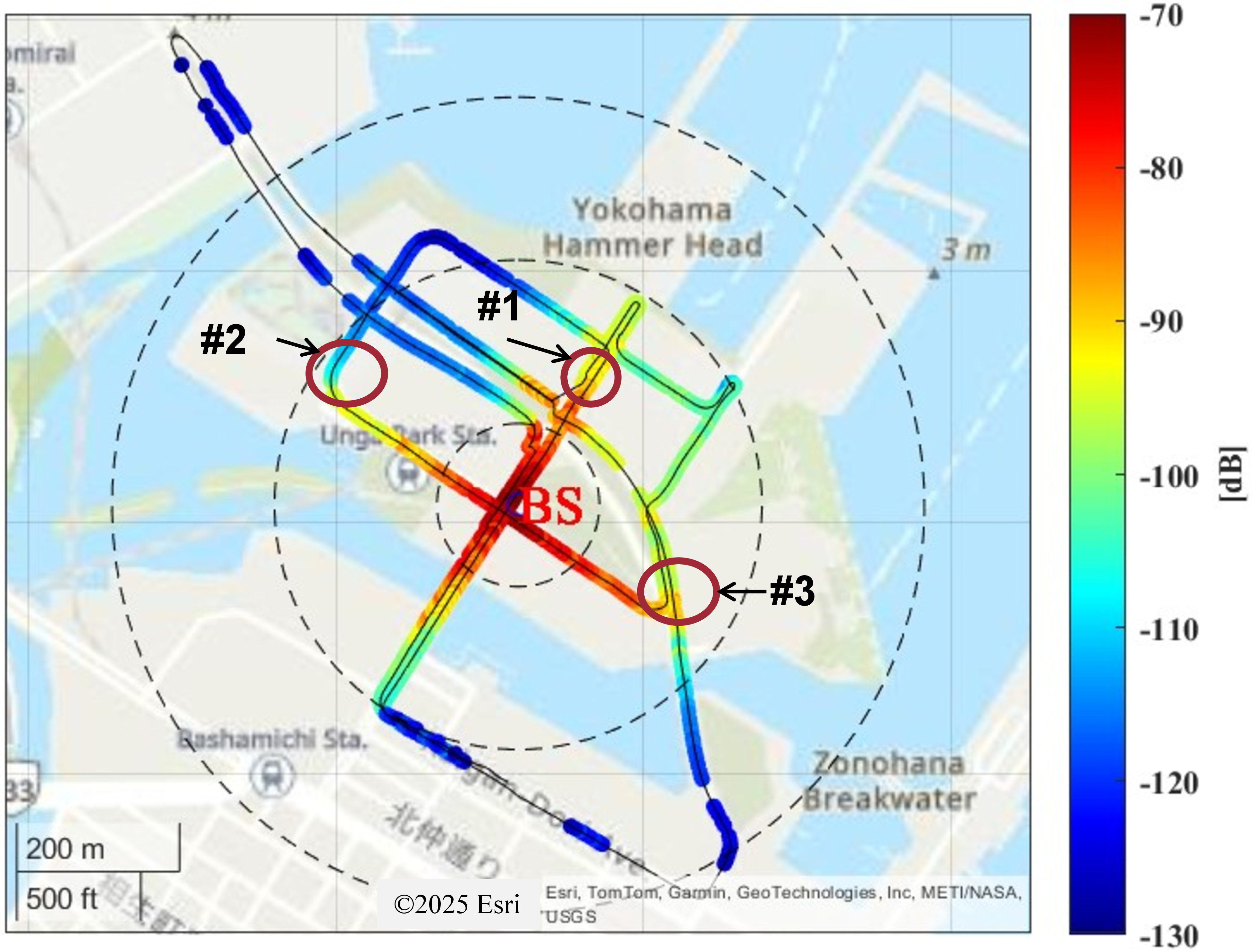}} 
\caption{Route-wise PL maps. The circled regions labeled \#1--\#3 denote representative receiver locations where the measured PL shows localized deviations from the main regression trend in Fig.~\ref{PL_fit}, typically near intersections, corners, or local visibility-transition regions.} \label{Routes_PL}
\end{figure*}

\begin{table*}[!t]
  \centering
  \caption{Path loss fitting results at $4.85$~GHz. 
  The CI and FI parameters are fitted to the measurements. 
  The ITU-R entries are scenario-level ABG reference parameters, with the site-general above-rooftop and below-rooftop models corresponding to UMa and UMi, respectively. 
  The 3GPP entries include the native TR~38.901~\cite{TR138-901} shadow-fading standard deviation $\sigma_{\rm ref}$ and equivalent parameters obtained by refitting the corresponding deterministic UMa/UMi LoS and NLoS mean PL curves over the same distance range. The equivalent 3GPP parameters are derived approximation parameters and should not be interpreted as native TR~38.901 coefficients.}
  \label{tab:pl-params}

  \begingroup
  \footnotesize
  \setlength{\tabcolsep}{2.6pt}
  \renewcommand{\arraystretch}{1.08}
  \begin{tabular}{c c c c c c c c}
    \toprule
    Route & Scen. & State
    & CI fit $(n,\sigma,\mathrm{RMSE})$
    & FI fit $(\alpha,\beta,\sigma,\mathrm{RMSE})$
    & ITU-R ABG $(\alpha,\beta,\gamma,\sigma)$
    & 3GPP~ref. ($\sigma$)
    & 3GPP eq. $(\alpha,\beta,\mathrm{RMSE})$ \\
    \midrule

    \multirow{2}{*}{{\tt Area1}} 
    & \multirow{2}{*}{UMa} 
    & LoS
      & $(2.31,\,3.70,\,3.66)$
      & $(1.98,\,52.59,\,3.67,\,3.64)$
      & $(2.29,\,28.60,\,1.96,\,3.48)$
      & $4.00$
      & $(2.26,\,40.52,\,0.46)$ \\

    & & NLoS
      & $(2.89,\,4.50,\,4.49)$
      & $(2.91,\,45.45,\,4.50,\,4.49)$
      & $(4.39,\,-6.27,\,2.30,\,6.89)$
      & $6.00$
      & $(3.90,\,27.37,\,0.09)$ \\
    \midrule
    \multirow{2}{*}{{\tt Area2}} 
    & \multirow{2}{*}{UMa} 
    & LoS
      & $(1.94,\,2.73,\,2.73)$
      & $(1.82,\,48.93,\,2.72,\,2.71)$
      & $(2.29,\,28.60,\,1.96,\,3.48)$
      & $4.00$
      & $(2.26,\,40.52,\,0.46)$ \\

    & & NLoS
      & $(2.90,\,3.74,\,3.73)$
      & $(1.21,\,90.13,\,3.38,\,3.37)$
      & $(4.39,\,-6.27,\,2.30,\,6.89)$
      & $6.00$
      & $(3.90,\,27.37,\,0.09)$ \\

    \midrule

    \multirow{2}{*}{{\tt Area3}} 
    & \multirow{2}{*}{UMi} 
    & LoS
      & $(2.04,\,4.41,\,4.41)$
      & $(2.55,\,35.26,\,4.10,\,4.10)$
      & $(2.07,\,31.23,\,2.06,\,4.91)$
      & $4.00$
      & $(2.84,\,31.37,\,2.10)$ \\

    & & NLoS
      & $(2.68,\,7.32,\,7.31)$
      & $(4.61,\,-1.01,\,6.66,\,6.66)$
      & $(3.73,\,16.02,\,2.26,\,7.62)$
      & $7.82$
      & $(3.53,\,37.01,\,0.00)$ \\

    \bottomrule
  \end{tabular}
  \vspace{1mm}
  \begin{flushleft}
  \footnotesize Note: $\sigma$ and RMSE are in dB, and the FI intercept $\beta$ is in dB. The ITU-R ABG parameters are native site-general reference parameters. 
  For 3GPP, $\sigma_{\rm ref}$ is the native TR~38.901 shadow-fading standard deviation for the corresponding UMa/UMi LoS/NLoS model. 
  The 3GPP mean-curve equivalent parameters are obtained by refitting the deterministic TR~38.901 mean PL curves and are included only as compact approximations of the 3GPP mean PL behavior. Their RMSE represents the 3GPP approximation error and should not be interpreted as shadow fading.
  \end{flushleft}
  \endgroup
\end{table*}
\begin{figure*}[t]
\centering
\subfigure[{\tt Area1} (UMa).\label{fig:DS_R3}]{\includegraphics[width=0.31\linewidth]{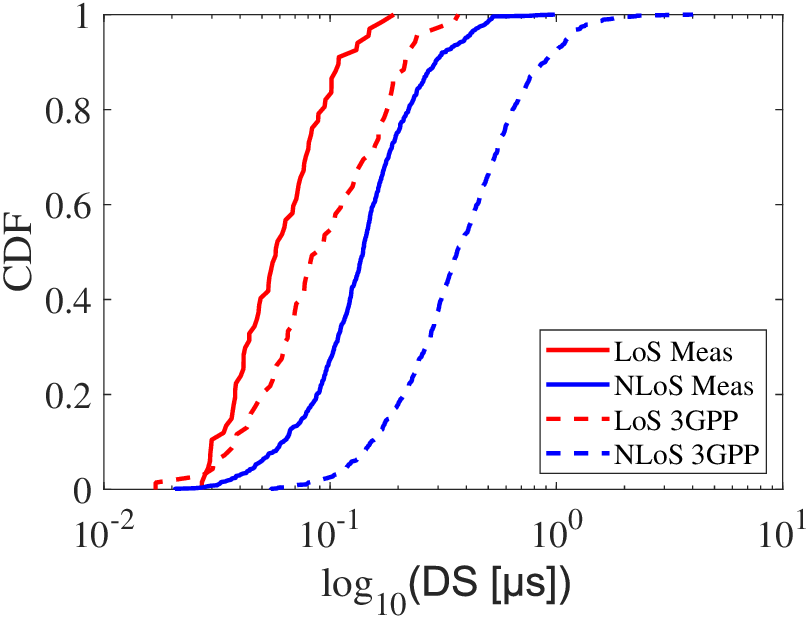}} 
\subfigure[{\tt Area2} (UMa).\label{fig:DS_R4}]{\includegraphics[width=0.31\linewidth]{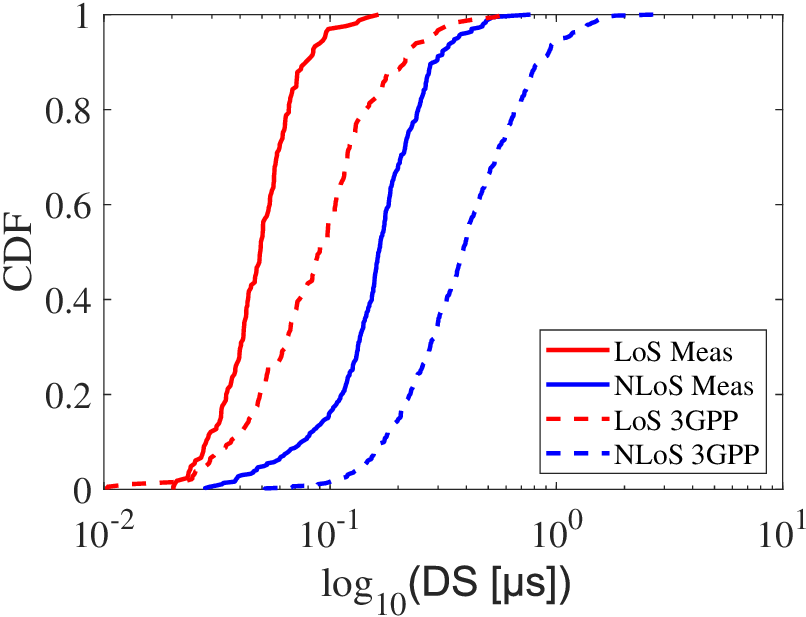}} 
\subfigure[{\tt Area3} (UMi).\label{fig:DS_R6}]{\includegraphics[width=0.31\linewidth]{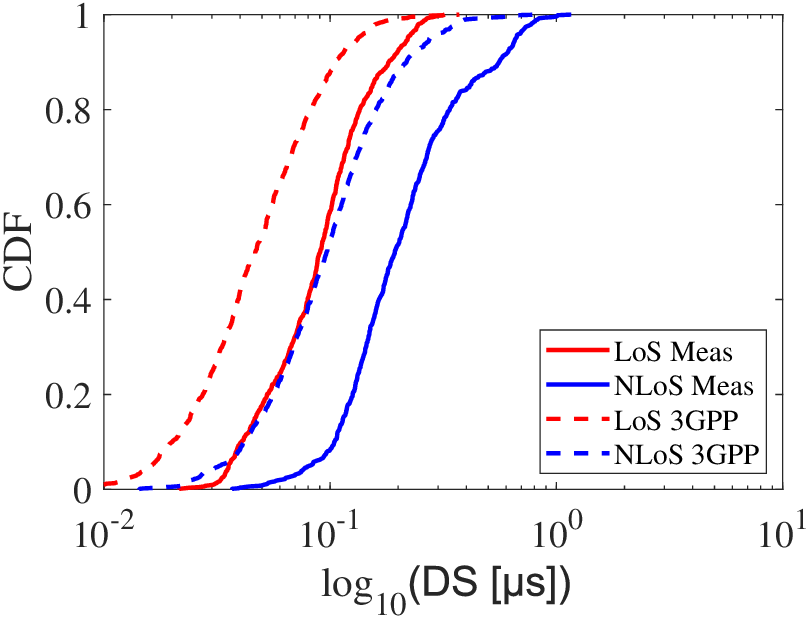}} 
\caption{Empirical CDFs of the RMS delay spread measured at $4.85$~GHz in the UMa and UMi routes, compared with the corresponding 3GPP reference distributions.
\label{fig:DS_all}}
\end{figure*}
\begin{figure*}[t]
\centering
\subfigure[{\tt Area1} (UMa).\label{fig:AS_R3_new}]
{\includegraphics[width=0.31\linewidth]{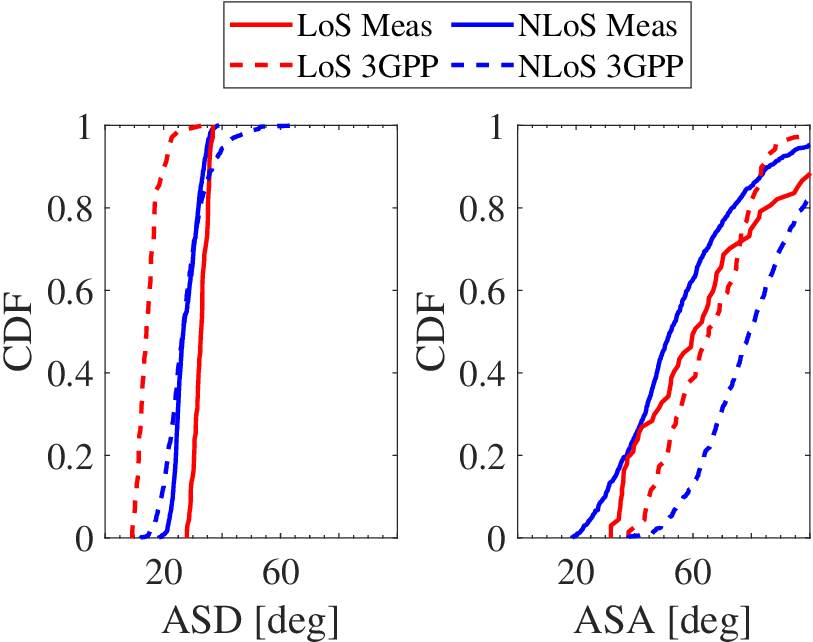}}
\subfigure[{\tt Area2} (UMa).\label{fig:AS_R4_new}]
{\includegraphics[width=0.31\linewidth]{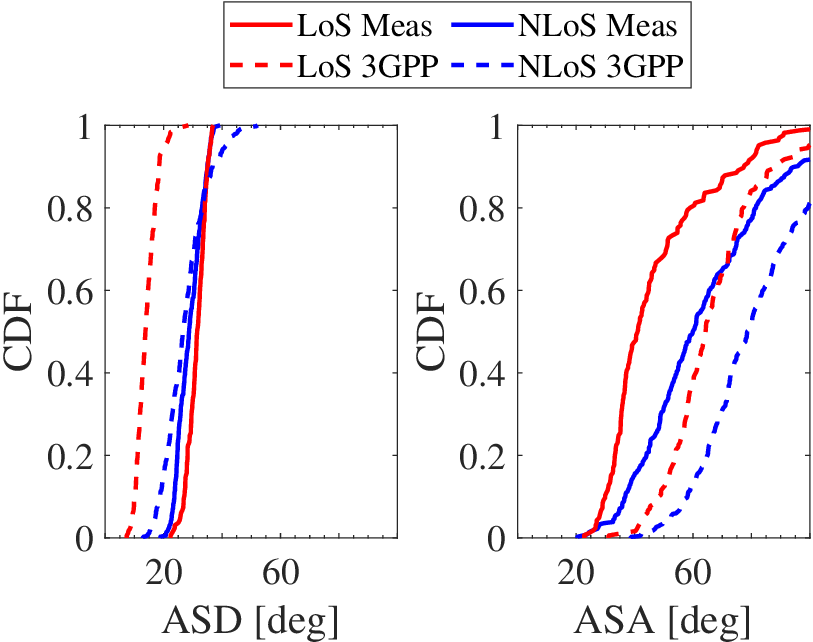}}
\subfigure[{\tt Area3} (UMi).\label{fig:AS_R6_new}]
{\includegraphics[width=0.31\linewidth]{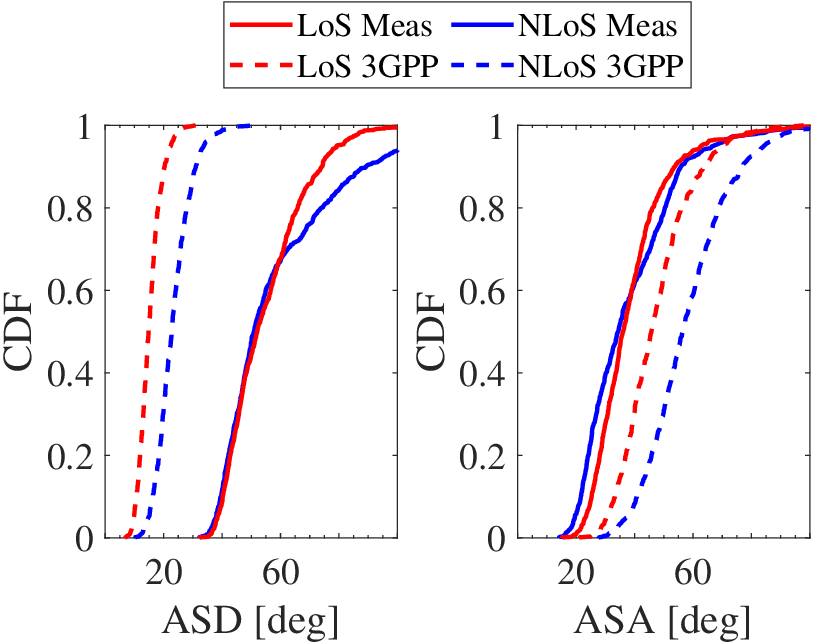}}

\caption{Empirical CDFs of the ASD and ASA measured at $4.85$~GHz in the UMa and UMi routes, together with the corresponding 3GPP reference distributions.}
\label{fig:AS}
\end{figure*}

\section{Channel Characterization}

\subsection{Path Loss Modeling}
To characterize the large-scale received-power behavior, the total received power at each snapshot was estimated by incoherently summing the powers of the extracted MPCs, which suppresses small-scale phase interference. The path loss (PL) is then computed as 
\begin{eqnarray}
\PL~\mathrm{[dB]} & = -10\log_{10}\sum_{l} |\hat{\gamma}_l|^2,
\label{eq:omniPL_MPC}
\end{eqnarray}
where $\hat{\gamma}_l$ denotes the estimated complex path weight of the $l$th extracted MPC. The resulting PL should therefore be interpreted as an MPC-based estimate of the large-scale received power. 
The PL is modeled using the close-in (CI) free-space reference-distance model with $d_0=1$~m and the floating-intercept (FI) model \cite{Rappaport_mmWave,[Park]CI_model}:
\begin{multline}
\PL^\CI (f_c, d_{\ThreeD}) ~ [\mathrm{dB}] = \mathrm{FSPL}(f_c,1\,\mathrm{m}) + \\ 10 \, n \, \log_{10}\left(\frac{d_{\ThreeD}}{d_0}\right) +\chi_\sigma,
\label{PL_CI}
\end{multline}
\begin{multline}
\PL^\FI (f_c, d_{\ThreeD})~ [\dB] = 10 \, \alpha \, \log_{10}\left({d_{\ThreeD}}\right)+ \beta + \chi_\sigma,
\label{PL_FI}
\end{multline}
where $\mathrm{FSPL}(f_c,1\,\mathrm{m})= 32.4+20\log_{10}\!\left(\frac{f_c}{1~\GHz}\right)$ denotes the free space path loss (FSPL) at $1.0$~m, $n$ is the path loss exponent (PLE), and $\chi_\sigma$ denotes the log-normal SF term \cite{sun_VTC2016, Yubei_Spectrum_Sandbox}. Furthermore, the ITU-R ABG model \cite{ITU-R_P.1411} was employed as a benchmark. It is expressed as
\begin{multline}
\PL^\ABG (f_c, d_{\ThreeD})~ [\dB] =  10 \, \alpha \, \log_{10}({d_{\ThreeD}}) + \beta + \\ 10 \, \gamma \, \log_{10}({f_c}) + \chi_\sigma,
\label{PL_ABG}
\end{multline}
where $\alpha$, $\beta$, and $\gamma$ are the distance slope, intercept, and frequency slope, respectively.

Fig.~\ref{PL_fit} and Table~\ref{tab:pl-params} summarize the measured PL and the corresponding CI/FI fits for the three routes. In LoS, the CI PL exponents remain close to free-space, with $n=2.31$, $1.94$, and $2.04$ for {\tt Area1}, {\tt Area2}, and {\tt Area3}, respectively, and moderate SF spreads of $3.70$~dB, $2.73$~dB, and $4.41$~dB. In NLoS, the CI exponents increase to $2.89$, $2.90$, and $2.68$, while the SF spread also becomes larger, most notably in {\tt Area3} NLoS with $\sigma=7.32$~dB. Table~\ref{tab:pl-params} reports RMSE values for the measured CI/FI fits and for the 3GPP-equivalent approximation. The measured-fit RMSE values closely follow the corresponding SF spreads; for example, the CI RMSE ranges from $2.73$ to $4.49$~dB in the UMa routes and increases to $7.31$~dB in the {\tt Area3} UMi NLoS case, confirming the stronger PL variability of the street-level NLoS route. The FI model provides only limited numerical improvement overall. In {\tt Area1}, the CI and FI fits are nearly identical. In {\tt Area2} NLoS, FI reduces the spread from $3.74$~dB to $3.38$~dB, but with $\alpha=1.21$ and $\beta=90.13$~dB, indicating reduced physical interpretability. In {\tt Area3}, FI gives moderate reductions in spread for both LoS ($\alpha=2.55$, $\beta=35.26$~dB) and NLoS ($\alpha=4.61$, $\beta=-1.01$~dB), although the floating intercept remains only a regression parameter.
For standardized-reference comparison, Table~\ref{tab:pl-params} includes both ITU-R and 3GPP PL references. The ITU-R entries are taken directly from the site-general ABG parameter sets in~\cite{ITU-R_P.1411}, using the above-rooftop category for UMa and the below-rooftop category for UMi. For 3GPP, the corresponding TR~38.901 UMa/UMi LoS and NLoS PL models are used~\cite{TR138-901}. The listed 3GPP $\sigma_{\rm ref}$ values are the native TR~38.901 shadow-fading standard deviations, whereas the 3GPP-equivalent $\alpha$, $\beta$, and RMSE values are obtained by refitting the deterministic 3GPP mean PL curves over the same distance range used for the comparison, following the type of equivalent-parameter extraction used in~\cite{5G_Models_100G, Poddar_2025}. These equivalent values are included only as compact approximations of the 3GPP mean PL behavior and should not be interpreted as native TR~38.901~\cite{TR138-901} coefficients. In particular, the 3GPP-equivalent RMSE represents deterministic curve-approximation error, not a shadow-fading standard deviation. 

Relative to the ITU-R and 3GPP standardized references, the measured LoS SF spreads are generally comparable to or lower than the reference values, except for {\tt Area3} LoS, where the measured CI spread is slightly above the 3GPP LoS reference value. In NLoS, the UMa routes show lower variability than the standardized reference levels, whereas the UMi route ({\tt Area3}) is closer to them. This comparison is intended only to place the measured route-specific PL behavior within established scenario-level references, not to validate or recalibrate either standard.
Overall, the results indicate near-free-space LoS behavior and stronger route-dependent variability in NLoS, with {\tt Area3} NLoS representing the most challenging case. The measured CI exponents are also consistent with nearby-frequency urban PL measurements: the LoS values, $n=1.94$--$2.31$, remain close to free-space behavior, while the NLoS values, $n=2.68$--$2.90$, fall within the higher range typically reported near $6.75$--$10.1$~GHz due to blockage and street-canyon propagation~\cite{3GPP-TR138_921, Miao_2024, Miao_Sub6, Miao_2025}. Thus, the present $4.85$~GHz PL exponents are not anomalous, although their exact values remain route- and geometry-dependent.
This bounded interpretation is consistent with recent multi-frequency PL studies~\cite{Poddar_2025}, which show that standardized model assessment should rely on aggregated evidence across frequencies and campaigns, and that CI/CIF-type formulations tend to remain more stable than FI/ABG-type parameterizations when frequency support is sparse.

Figure~\ref{Routes_PL} shows the measured PL values projected onto the corresponding satellite maps, where the color-coded trajectories represent the PL variation along each measurement route. Rx locations with insufficient post-processing SNR or duplicate/stationary samples caused by vehicle stops were excluded to retain only valid samples for continuous PL characterization.
Localized deviations from the main PL trend are observed near intersections, corners, and LoS--NLoS transition regions, as highlighted by \#1--\#3 in Figs.~\ref{PL_fit} and~\ref{Routes_PL}. In these regions, reflected and diffracted components can temporarily increase the received power relative to the surrounding NLoS trend, producing scatter that is captured by the stochastic $\chi_\sigma$ term in the CI/FI models. In the UMi route, the low street-level BS height further enhances ground and building-facade interactions, which can lead to slightly higher received power than the free-space reference at some LoS distances due to the constructive combination of the direct and reflected components.
\begin{table*}[!t]
  \scriptsize
  \caption{Model parameters obtained from measurements at $4.85$~GHz (Mean $\mu$ and Std. Dev. $\sigma$).}
  \label{table_param_tab_LoS}
  \centering
  \begin{tabular}{lccccccccccc}
    \toprule
    \multirow{2}{*}{Parameter} & \multirow{2}{*}{}
      & \multicolumn{2}{c}{{\tt Area1} (UMa)} 
      & \multicolumn{2}{c}{{\tt Area2} (UMa)} 
      & \multicolumn{2}{c}{3GPP (UMa)} 
      & \multicolumn{2}{c}{{\tt Area3} (UMi)} 
      & \multicolumn{2}{c}{3GPP (UMi)} \\
    \cmidrule(lr){3-4} \cmidrule(lr){5-6} \cmidrule(lr){7-8} \cmidrule(lr){9-10} \cmidrule(lr){11-12}
      & & LoS & NLoS & LoS & NLoS & LoS & NLoS & LoS & NLoS & LoS & NLoS \\
    \midrule

    \multirow{2}{*}{\shortstack[l]{RMS DS\\$\log_{10}(\mathrm{DS}/1\,\mathrm{s})$}}
      & $\mu$    & $-6.9931$ & $-6.8815$ & $-6.9953$ & $-6.9443$ & $-6.9576$ & $-7.0022$ & $-6.9338$ & $-6.8112$ & $-6.4813$ & $-6.6328$ \\
      & $\sigma$ & $0.18$ & $0.28$ & $0.35$ & $0.31$ & $0.66$ & $0.70$ & $0.28$ & $0.33$ & $0.39$ & $0.32$ \\
    \midrule
    
    \multirow{2}{*}{\shortstack[l]{RMS ASD \\ $\log_{10}(\mathrm{ASD}/1^\circ)$}}
      & $\mu$   & $1.5139$ & $1.4392$ & $1.4928$ & $1.4570$ & $1.0786$ & $1.4851$ & $1.7280$ & $1.7444$ & $1.2809$ & $1.2513$ \\
      & $\sigma$ & $0.03$ & $0.06$ & $0.05$ & $0.06$ & $0.30$ & $0.45$ & $0.13$ & $0.14$ & $0.23$ & $0.34$ \\
    \midrule
    
    \multirow{2}{*}{\shortstack[l]{RMS ASA \\ $\log_{10}(\mathrm{ASA}/1^\circ)$}}
      & $\mu$    & $1.7773$ & $1.7149$ & $1.6456$ & $1.7755$ & $1.8100$ & $2.0298$ & $1.5652$ & $1.5449$ & $1.2211$ & $1.4213$ \\
      & $\sigma$ & $0.16$ & $0.17$ & $0.15$ & $0.16$ & $0.20$ & $0.31$ & $0.14$ & $0.17$ & $0.16$ & $0.26$ \\
    \midrule
    
    \multirow{2}{*}{$K$-factor (dB)}
      & $\mu$    & $-0.82$ & \NA & $2.51$ & \NA & $9.00$ & \NA & $0.69$ & \NA & $9$ & \NA \\
      & $\sigma$ & $2.72$ & \NA & $3.00$ & \NA & $3.5$ & \NA & $3.01$ & \NA & $5$ & \NA \\
    \bottomrule
  \end{tabular}
\end{table*}
\begin{figure}[t]
\centering
\includegraphics[width=0.9\linewidth]{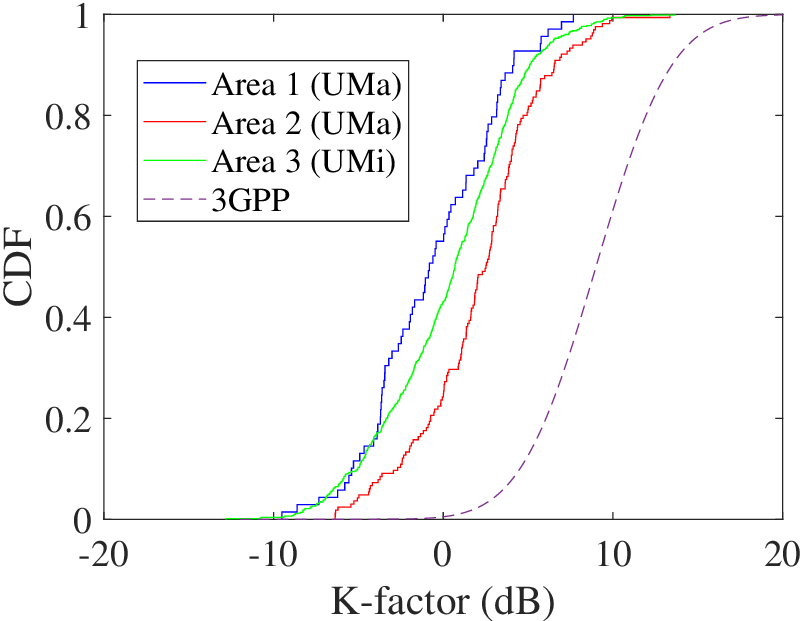}
\caption{Empirical CDFs of the wideband snapshot-based Rician $K$-factor estimated from the reconstructed $4.85$~GHz channels in the measured UMa and UMi routes, compared with the corresponding 3GPP reference distribution} 
\label{fig:K-factor}
\end{figure}

\subsection{Large-Scale Parameters Extraction}

The large-scale parameters (LSPs), including the root-mean-square (RMS) delay spread (DS), azimuth spread of departure (ASD), and azimuth spread of arrival (ASA), were computed from the extracted MPC parameters $\hat{\gamma}_{l}$, $\hat{\tau}_{l}$, $\hat{\varphi}_{\T,l}$, and $\hat{\varphi}_{\R,l}$, representing the complex path weight, delay, angle of departure, and angle of arrival of the $l$th MPC, respectively. Table~\ref{table_param_tab_LoS} summarizes the resulting route-wise LSP statistics at $4.85$~GHz and the corresponding 3GPP reference parameters used for comparison.

\vspace{2mm}
\subsubsection{Delay Spread}
The RMS DS quantifies the temporal dispersion, computed using the standard power-weighted second central moment definition \cite{TR138-901}:
\begin{equation}
  \tau_{\mathrm{DS}}  =  \sqrt{\frac{\sum_l \hat{\tau}_l^{2} \left|\hat{\gamma}_{l}\right|^{2}}{\sum_l\left|\hat{\gamma}_{l}\right|^{2}} - \left(\frac{\sum_l\hat{\tau}_l\left|\hat{\gamma}_{l}\right|^{2}}{\sum_l\left|\hat{\gamma}_{l}\right|^{2}}\right)^{2}}.
  \label{eqn:ds}
\end{equation}
The DS empirical cumulative distribution functions (eCDFs) in Fig.~\ref{fig:DS_all} are compared against the corresponding 3GPP UMa/UMi reference distributions. In the two UMa routes ({\tt Area1} and {\tt Area2}), the measured DS distributions are generally smaller than the 3GPP references under both LoS and NLoS, indicating shorter excess-delay behavior in these particular measurements. In the UMi route ({\tt Area3}), the measured DS is generally larger than the 3GPP UMi reference, especially under NLoS, which is plausibly associated with stronger street-canyon confinement and persistent local reflections along the measured route.

These observations should be interpreted only in the context of the $4.85$~GHz Yokohama measurement campaign. Since the 3GPP model is a generic stochastic parameterization derived from multiple campaigns across sites and frequency bands, the comparison is not intended to imply that 3GPP generally overestimates or underestimates DS, but only that the default scenario-level assumptions do not fully reproduce the DS behavior observed in this particular measurement campaign. Accordingly, the comparison is reported as campaign-specific context, not as evidence of a general deficiency of the 3GPP model.

\vspace{2mm}
\subsubsection{Angular Spreads}
The azimuth angular spread is computed using the standard circular power-weighted definition commonly used in channel modeling \cite{TR138-901, WINNER2007}:
\begin{equation}
  \varphi_{\mathrm{AS},x}  =  \sqrt{-2\ln \left|\frac{\sum_l \exp\!\left(\j\hat{\varphi}_{x,l}\right)\left|\hat{\gamma}_{l}\right|^{2}}{\sum_l\left|\hat{\gamma}_{l}\right|^{2}}\right|},
  \label{eqn:as}
\end{equation}
where the subscript $x=$~`T' corresponds to ASD and $x=$~`R' corresponds to ASA, consistent with the earlier definitions. 

The measured ASA/ASD distributions in Fig.~\ref{fig:AS} differ from the corresponding 3GPP references in all three routes. In the measured UMa routes, the ASD tends to be broader than the corresponding reference, while the ASA is generally narrower, indicating that the departure-side and arrival-side angular dispersions observed in this campaign differ from those represented by the generic stochastic model. A similar pattern is also observed in the UMi route, where the measured ASD remains comparatively large, whereas the ASA is more confined. These observations should be interpreted only in the context of the present $4.85$~GHz Yokohama measurement campaign. Since the 3GPP model is a generic stochastic parameterization derived from multiple campaigns across sites and frequency bands, this comparison is intended only as campaign-specific context and not as evidence of a general deficiency of the standardized angular-spread assumptions.

\begin{figure*}[t]
\centering
\subfigure[{\tt Area1} (UMa).\label{fig:Spread_R3}]{\includegraphics[width=0.31\linewidth]{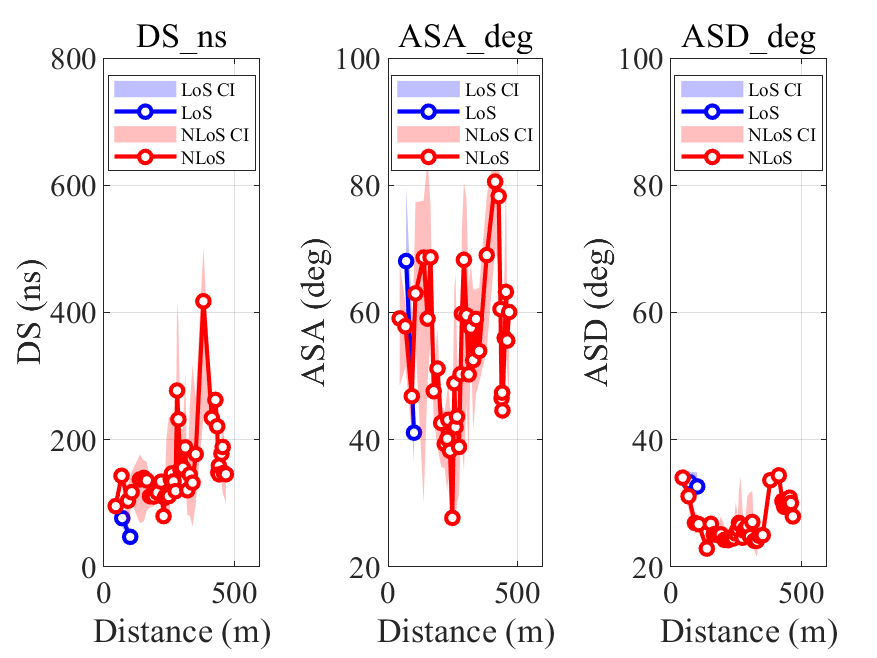}} 
\subfigure[{\tt Area2} (UMa).\label{fig:Spread_R4}]{\includegraphics[width=0.31\linewidth]{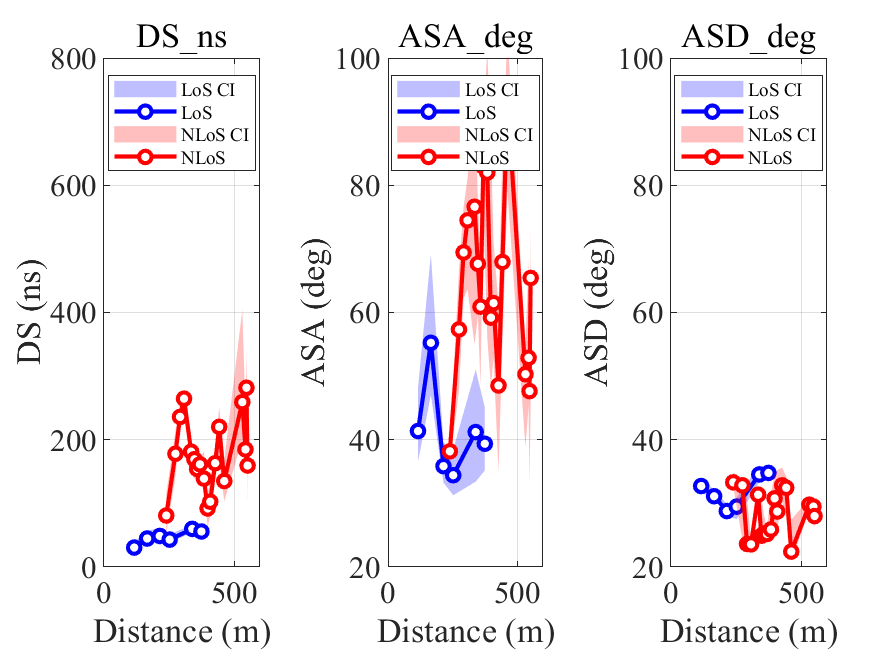}} 
\subfigure[{\tt Area3} (UMi).\label{fig:Spread_R6}]{\includegraphics[width=0.31\linewidth]{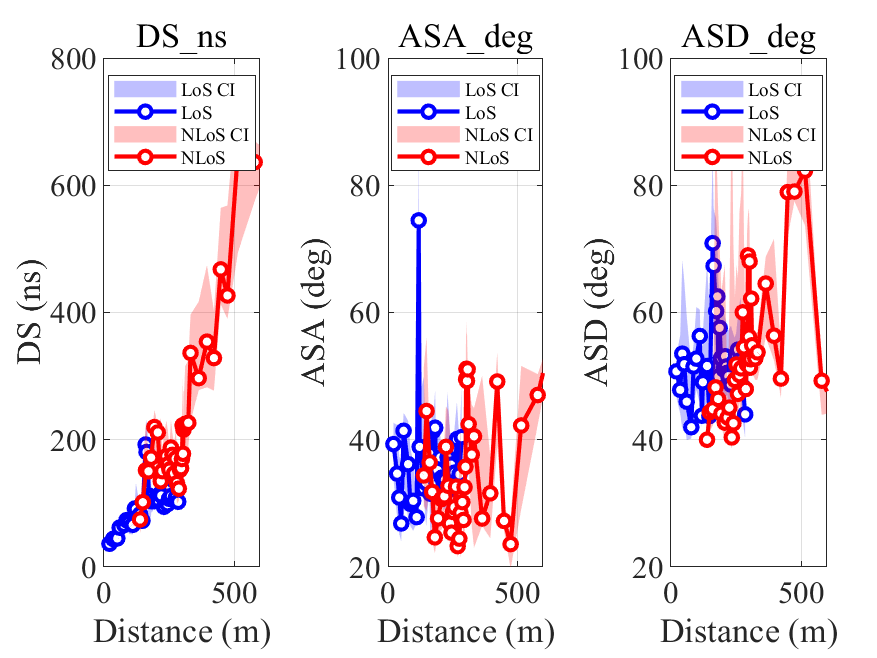}} 

\caption{LSP distance dependency. LoS/NLoS median curves with shaded $5$-$95\%$ CI bands for DS, ASA, and ASD versus distance.
\label{fig:Spread}}
\end{figure*}

\vspace{2mm}
\subsubsection{Rician $K$-factor}
The Rician $K$-factor classically quantifies the ratio between the dominant specular component and the diffuse multipath in a narrowband temporal-fading process. In this work, however, it is estimated from a single reconstructed wideband channel snapshot by applying the method-of-moments estimator in \cite[Eq.~(8)--(10)]{Tang_Estimating_K-factor} to the reconstructed CTF $H(f)$. Let $H_i$ denote the complex channel coefficient at the $i$th frequency sample. The first and second sample moments are
\begin{align}
  G_a &= \frac{1}{N}\sum_{i=1}^{N} \bigl|H_i\bigr|^2,  \\
  G_v &= \frac{1}{N-1} \left(\sum_{i=1}^{N} \bigl|H_i\bigr|^4 - N G_a^2\right),
  \label{Eq:Moments}
\end{align}
where $N$ is the number of frequency tones. The corresponding moment-based estimate is
\begin{equation}
  \hat{K} = \frac{\sqrt{G_a^2 - G_v}}{\,G_a - \sqrt{G_a^2 - G_v}\,}.
  \label{Eq:K-factor}
\end{equation}
The estimator in \cite{Tang_Estimating_K-factor} uses the second- and fourth-order moments of the reconstructed frequency-domain channel samples within a single wideband snapshot to derive a compact indicator of specular-to-diffuse dominance. In the present application, this provides a useful descriptive proxy, but it is not strictly equivalent to the classical narrowband temporal Rician $K$-factor used in standardized stochastic models. In particular, the estimate can be influenced by the frequency selectivity of the channel over the analyzed bandwidth and should therefore be interpreted as a snapshot-based wideband proxy rather than as a directly comparable 3GPP $K$-factor quantity.

Fig.~\ref{fig:K-factor} compares the eCDFs of this proxy metric across the three areas with the 3GPP reference distribution \cite{TR138-901}. The estimated proxy values are generally lower than the 3GPP reference in all three routes. However, this should not be interpreted as a direct physical discrepancy, since the comparison involves different definitions and estimation domains: the values in this work are wideband snapshot-based proxy estimates, whereas the 3GPP parameters are scenario-level, stochastic, narrowband quantities. Accordingly, Fig.~\ref{fig:K-factor} is included only as a secondary descriptive result, not as a primary basis for model conclusions.

\subsection{Distance-Dependent LSP Trends}
Using the per-snapshot DS, ASA, and ASD values, we next examine how the measured dispersion evolves along each route. Distance-dependent analysis is commonly used in large-scale propagation studies to characterize how channel behavior changes with Tx--Rx separation, particularly for PL and large-scale fading trends~\cite{sun_distance2016}. In the present work, this idea is extended descriptively to the measured LSPs to examine route-wise variations in delay and angular dispersion. This is relevant to geometry-based stochastic channel models (GSCMs) such as WINNER II, QuaDRiGa, and 3GPP-type spatially consistent channel models \cite{WINNER2007, Quadriga_3DMulticell, TR138-901}, which generally represent LSPs through scenario-level statistical distributions together with decorrelation distances or related spatial-consistency procedures. The present analysis is therefore intended to provide empirical route-wise observations of how the measured LSPs evolve with distance, and to illustrate aspects of local channel evolution that a purely scenario-level analysis may not fully capture.

For each route $\mathcal{R}\in\{\mbox{{\tt Area1}, {\tt Area2}, {\tt Area3}}\}$ and propagation state $\mathcal{S}\in\{\mathrm{LoS,NLoS}\}$, the valid snapshots are first sorted according to their 3D Tx--Rx separation $\{d_n\}_{n=1}^{N}$. They are then partitioned into disjoint adaptive bins with at least $N_{\min}=20$ samples and a maximum bin width of $\Delta d_{\max}=50$~m. This produces narrower bins in densely sampled portions of the route and wider bins in sparse segments. The representative distance of the $k$th bin is taken as the bin center,
\begin{eqnarray}
\tilde{d}_k = \frac{d_{k-1} + d_k}{2}.
\end{eqnarray}
For each bin and each parameter $X\in\{\DS,\ASA,\ASD\}$, the sample median is used as a robust summary of the local typical behavior. To quantify the uncertainty of this median estimate, $5$--$95\%$ bootstrap intervals are computed using $B=1000$ resamples with replacement from the samples in each bin. These intervals, therefore, describe the uncertainty of the estimated median, rather than the full physical spread of the channel values within the bin.
\begin{figure*}[t]
\centering

\subfigure[{\tt Area1} (UMa).\label{fig:SC_R3}]{\includegraphics[width=0.31\linewidth]{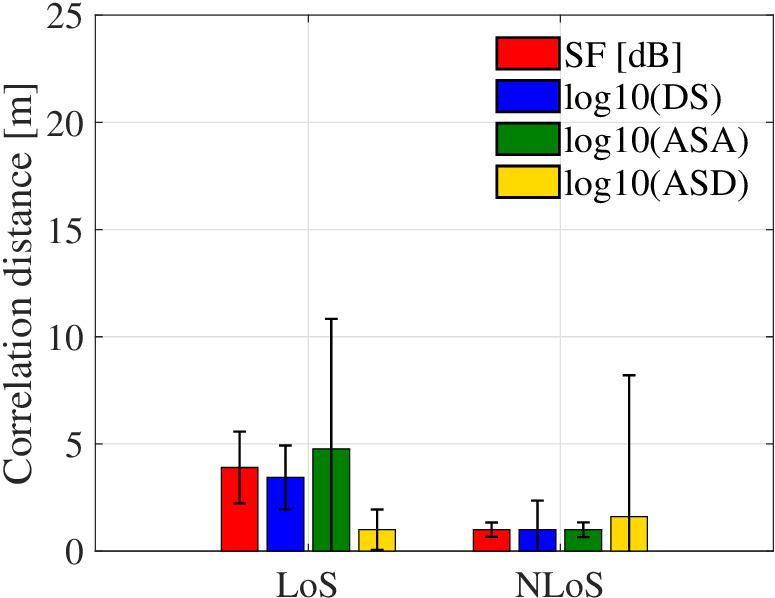}} 
\subfigure[{\tt Area2} (UMa).\label{fig:SC_R4}]{\includegraphics[width=0.31\linewidth]{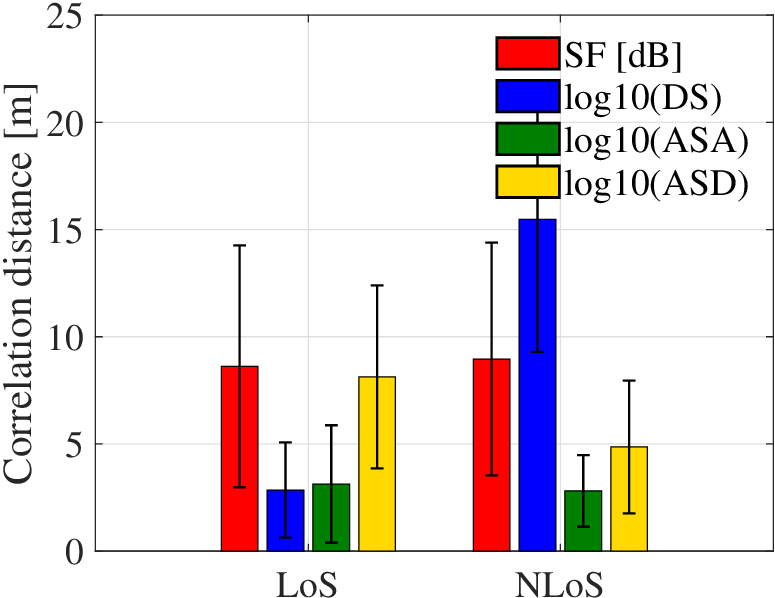}} 
\subfigure[{\tt Area3} (UMi).\label{fig:SC_R6}]{\includegraphics[width=0.31\linewidth]{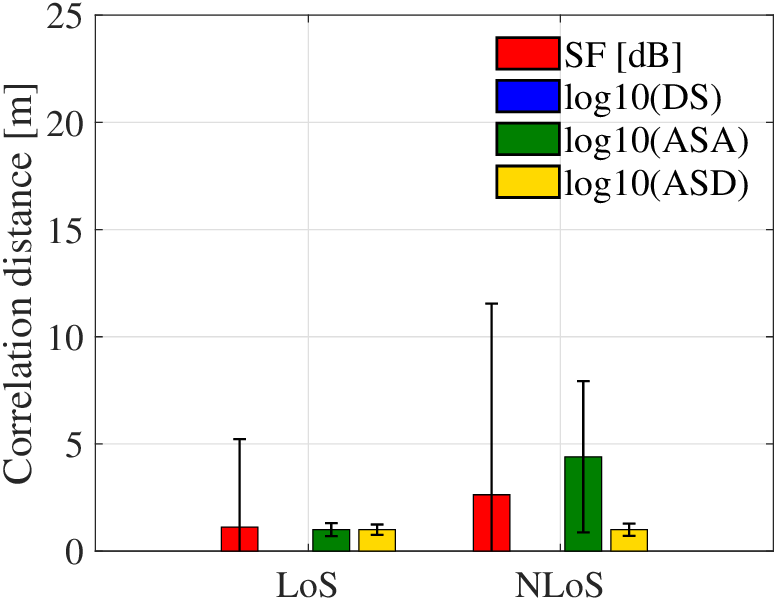}}  

\caption{Spatial consistency. Bars show the estimated decorrelation distance. The vertical error bars denote the $95\%$ confidence intervals of the decorrelation distance.}
\label{fig:SC}
\end{figure*}
Fig.~\ref{fig:Spread} shows the resulting distance-dependent median trends for LoS and NLoS segments in the three routes. In the UMa routes, DS is generally small to moderate in LoS and larger in NLoS, whereas ASA shows substantial route-to-route and state-dependent variability. ASD remains comparatively confined in both UMa routes. In the UMi route ({\tt Area3}), the NLoS DS exhibits a much stronger increase with distance, reaching substantially larger values than in the UMa cases, while both ASA and ASD also show stronger local variability.

These results indicate that the measured LSPs can exhibit pronounced route dependence and non-stationary evolution, particularly in UMi NLoS. This is relevant for simulations because LSP evolution directly affects how the channel changes along a user trajectory, including the smoothness of delay- and angle-related channel dynamics. If such local distance dependence is not represented, scenario-level models may still reproduce average statistics. Still, they may not fully capture the measured route-wise spatial consistency and local environmental transitions. Accordingly, the trends shown here are intended as empirical summaries of the current measurement routes, not as universal laws or evidence of global stationarity. They nevertheless provide useful context for interpreting spatial-consistency assumptions in GSCM-based simulations in a local, route-dependent sense.
\begin{table*}[!t]
  \centering
  \caption{Measured $4.85$~GHz statistics and selected scenario-matched literature anchor points compiled from independent campaigns for the literature-referenced cross-band trend analysis.}
  \label{tab:Lit_Anchors}
  \renewcommand{\arraystretch}{1.2}
  \begin{tabular}{rcccccccc}
    \toprule
    \multirow{2}{*}{\textbf{Freq.}} & \multirow{2}{*}{\textbf{Ref.}} & \multirow{2}{*}{\textbf{Scenario}} & \multicolumn{2}{c}{\textbf{DS} (ns)} & \multicolumn{2}{c}{\textbf{ASA} ($^\circ$)} & \multicolumn{2}{c}{\textbf{ASD} ($^\circ$)} \\
    \cmidrule(lr){4-5} \cmidrule(lr){6-7} \cmidrule(lr){8-9}
     & & & LoS & NLoS & LoS & NLoS & LoS & NLoS \\
    \midrule
    
     $4.85$ GHz & {\tt Area1} & UMa & 101.60 & 131.37 & 59.88 & 51.87 & 32.65 & 27.49 \\
     $4.85$ GHz & {\tt Area2} & UMa & 101.09 & 113.68 & 44.22 & 59.64 & 31.10 & 28.64 \\
    
    $6$ GHz  & \cite{Miao_2024} & UMa & $38.90$ & $72.44$ & $30.90$ & $44.67$ & $7.1$ & $10.23$ \\ 
    $6.5$ GHz  & \cite{Poddar_2026} & UMa & $47.86$ & $97.72$ & $46.77$ & $52.48$ & $6.61$ & $18.2$ \\
    $7$ GHz & \cite{Miao_Sub6} & UMa & \NA & $24.5$ & \NA & \NA & \NA & \NA \\
    $8$ GHz  & \cite{Poddar_2026} & UMa & \NA & $127.44$ & \NA & $51.88$ & \NA & $3.98$ \\
    $13$ GHz  & \cite{Poddar_2026} & UMa & $44.25$ & $133.12$ & $37.05$ & $56.66$ & $13.76$ & $17.38$ \\
    $15$ GHz & \cite{Poddar_2026} & UMa & $16.6$ & $78.95$ & $32.36$ & $48.36$ & $14.79$ & $21.88$ \\
    $24.15$ GHz & \cite{Tsukada_TVT} & UMa & $107.47$ & \NA & $41.25$ & \NA & $28.82$ & \NA \\ \hline

    $4.85$ GHz & {\tt Area3} & UMi & 116.47 & 154.45 & 36.75 & 35.07 & 53.45 & 55.52 \\

    $6.5$ GHz  & \cite{Poddar_2026} & UMi & $6.92$ & \NA & $14.13$ & \NA & $10.96$ & \NA \\
    $6.75$ GHz  & \cite{NYU_FR1_FR3} & UMi & $62.8$ & $75.6$ & $21.4$ & $33.6$ & $20.7$ & $48.0$ \\ 
    $7$ GHz & \cite{Miao_Sub6} & UMi & $37.4$ & $67.13$ & $42$ & $43.3$ & \NA & \NA \\
    $8$ GHz  & \cite{Poddar_2026} & UMi & $22.39$ & $161.77$ & $41.69$ & $31.43$ & \NA & $7.59$ \\
    $10$ GHz  & \cite{Poddar_2026} & UMi & $25.54$ & $63.89$ & $21.84$ & $32.56$ & $10.96$ & $13.8$ \\
    $10.1$ GHz  & \cite{Poddar_2026} & UMi & $19.95$ & $38.9$ & $29.51$ & $58.88$ & $12.02$ & $17.38$ \\
    $11$ GHz & \cite{IEICE_Kim} & UMi & $6.3$ & $39.80$ & $63.1$ & $63.1$ & $20.0$ & $25.1$ \\ 
    $13$ GHz  & \cite{Poddar_2026} & UMi & $23.92$ & $66.38$ & $30.58$ & $47.22$ & $12.86$ & $20.42$ \\
    $13.5$ GHz  & \cite{Poddar_2026} & UMi & $6.46$ & \NA & $12.88$ & \NA & $9.55$ & \NA \\
    $15$ GHz & \cite{Miao_Sub6} & UMi & $24.55$ & $93.3$ & $9.33$ & $40.74$ & \NA & \NA \\
    $16.95$ GHz & \cite{NYU_FR1_FR3} & UMi & $46.5$ & $65.8$ & $15.3$ & $24.0$ & $24.0$ & $32.5$ \\
    $24.15$ GHz & \cite{Tsukada_TVT} & UMi & $65.48$ & \NA & $28.93$ & \NA & $33.78$ & \NA \\      
    $28.00$ GHz & \cite{NYU_OpenSquares}  & UMi & $20$ & $52$ & $14$ & $30$ & \NA    & \NA    \\
    \bottomrule
  \end{tabular}
\end{table*}

\subsection{Spatial Consistency of LSPs}
Spatial consistency describes how large-scale channel properties evolve smoothly as a user moves along a route \cite{Ademaj_GBSM, Suzuki_SC}. In practice, this is relevant to mobility-related evaluations such as handover, beam tracking, and spatially consistent stochastic simulation \cite{Aalborg_WCNC-Analysis, Poddar_3GPP-Rel19}. Standardized models such as \cite{ITU-R_P.1411} specify decorrelation distances at the scenario level. In contrast, the present results provide route-specific empirical observations of decorrelation behavior within the measured environment.

To characterize the spatial persistence of the measured LSPs, we analyze the spatial correlation of DS, ASA, ASD, and SF along the measurement routes. For the spatial-consistency analysis, SF is defined as the residual between the measured PL and the fitted large-scale PL model for the corresponding route and propagation condition. Therefore, the reported SF decorrelation distance characterizes the spatial persistence of the random large-scale fluctuation after removal of the deterministic distance-dependent PL trend. Given a spatially sampled LSP trace $X_k=X(k\Delta s)$ with an arc-length sampling interval $\Delta s$, the centered sequence is defined as $Y_k=X_k-\mu_X$. The normalized empirical autocorrelation function (ACF) is used here as a standard measurement-based estimator of spatial consistency:
\begin{equation}
\hat{R}_{X,\ell} \triangleq \hat{R}_{X}(\ell \Delta s) = \frac{\sum_{k} Y_k Y_{k+\ell}}{\sum_{k} Y_k^2},
\label{eq:emp_acf}
\end{equation}
where $\ell$ denotes the lag index. Although this formulation is motivated by a WSS-style autocorrelation calculation, several of the measured LSP traces exhibit distance-dependent evolution. Therefore, the resulting decorrelation distances should be interpreted as empirical local measures of spatial persistence along the route, rather than as evidence of strict wide-sense stationarity over the full route. Following the common use of exponential distance-dependent decorrelation models in standardized channel modeling \cite{ITU-R_P.1411,TR138-901}, the spatial decay is approximated as
\begin{equation}
R(\Delta d) \approx e^{-\Delta d/D_{\mathrm{corr}}},
\end{equation}
where $D_{\mathrm{corr}}$ denotes the decorrelation distance. It is estimated via nonlinear least-squares fitting to the empirical ACF:
\begin{equation}
\hat{D}_{\mathrm{corr}} = \arg\min_{D>0} \sum_{\ell} \left( \hat{R}_{X,\ell} - e^{-\ell \Delta s / D} \right)^2.
\label{eq:lsq_fit}
\end{equation}
The initial value is taken from the first approximate $1/e$ crossing of the empirical ACF. Estimation uncertainty is quantified using circular block bootstrapping. Specifically, resampled traces are constructed by concatenating randomly drawn index blocks with boundary wrap-around, and the decorrelation distance is re-estimated for each bootstrap realization.

Fig.~\ref{fig:SC} summarizes the estimated decorrelation distances for SF, $\log_{10}(\mathrm{DS}/1\,\mathrm{s})$, $\log_{10}(\mathrm{ASA}/1^\circ)$, and $\log_{10}(\mathrm{ASD}/1^\circ)$ in the three measured routes. Overall, the estimated distances are short, on the order of only a few meters, indicating limited spatial memory for the measured LSPs along the present routes. The relative behavior nevertheless differs across route and parameter type. 

In {\tt Area1} (UMa), all three parameters decorrelate rapidly under both LoS and NLoS. Under LoS, SF, DS, and ASA are of comparable magnitude, whereas ASD is somewhat smaller. Under NLoS, the estimated decorrelation distances remain short across all four quantities, with only a modest increase in ASD compared with the other quantities. The confidence intervals are generally moderate, indicating that the conclusion of short-range decorrelation in {\tt Area1} is reasonably stable even though the exact ordering among parameters is less certain. 

In {\tt Area2} (UMa), the decorrelation distances are clearly larger than in {\tt Area1}, especially under NLoS. The most prominent feature is the substantially larger decorrelation distance of DS in NLoS, which exceeds those of SF, ASA, and ASD and is accompanied by a relatively wide confidence interval. This suggests that, along this route, the delay-domain dispersion evolves more gradually than the angular structure, although the uncertainty also indicates sensitivity to local route nonstationarity. Under LoS in {\tt Area2}, SF and ASD also show noticeably larger persistence than the angular spreads in {\tt Area1}, whereas ASA remains comparatively short-range. Thus, {\tt Area2} is the clearest example in the dataset of route-dependent spatial persistence extending beyond only a few meters. 

In {\tt Area3} (UMi), the estimated decorrelation distances return to relatively short values. SF, ASA, and ASD remain confined to only a few meters in both LoS and NLoS, with relatively narrow confidence intervals for most cases, indicating limited but reasonably stable spatial persistence. For DS, the estimated autocorrelation does not reach the $1/e$ threshold within the adopted analysis window for either LoS or NLoS. Therefore, no strong conclusion can be drawn regarding the DS spatial persistence in {\tt Area3}.

Among the three routes, {\tt Area2} exhibits the strongest spatial persistence, particularly for NLoS DS, whereas {\tt Area1} and {\tt Area3} are dominated by short-range decorrelation for all reliably estimated parameters. The bootstrap confidence intervals further show that the robustness of the estimates depends strongly on route and parameter. Narrower intervals, as seen for several quantities in {\tt Area1} and {\tt Area3}, indicate that the available samples relatively well constrain the fitted decorrelation distance. In contrast, the wider intervals in {\tt Area2}, especially for NLoS DS, indicate that although the nominal decorrelation distance is larger, its exact value is less certain due to stronger local nonstationarity and limited effective sample support. Accordingly, these results are best interpreted as route-dependent empirical indicators of local spatial consistency within the measured environment. They provide useful measurement-based context for mobility-oriented simulations, but they are not intended as site-specific validations of, or replacements for, standardized scenario-level decorrelation models.

\begin{figure*}[t]
\centering

\subfigure[UMa.\label{fig:FC_UMa}]{\includegraphics[width=0.48\linewidth]{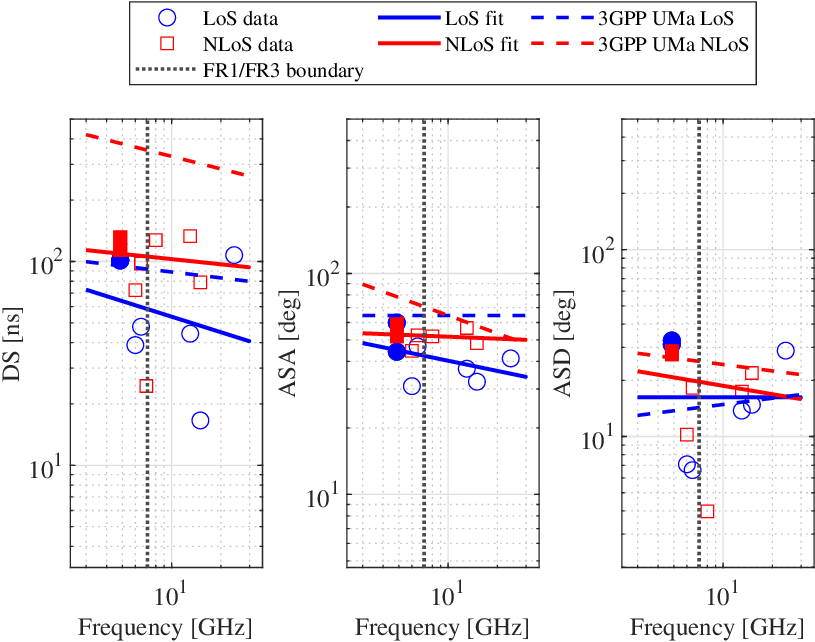}} 
\subfigure[UMi.\label{fig:FC_UMi}]{\includegraphics[width=0.48\linewidth]{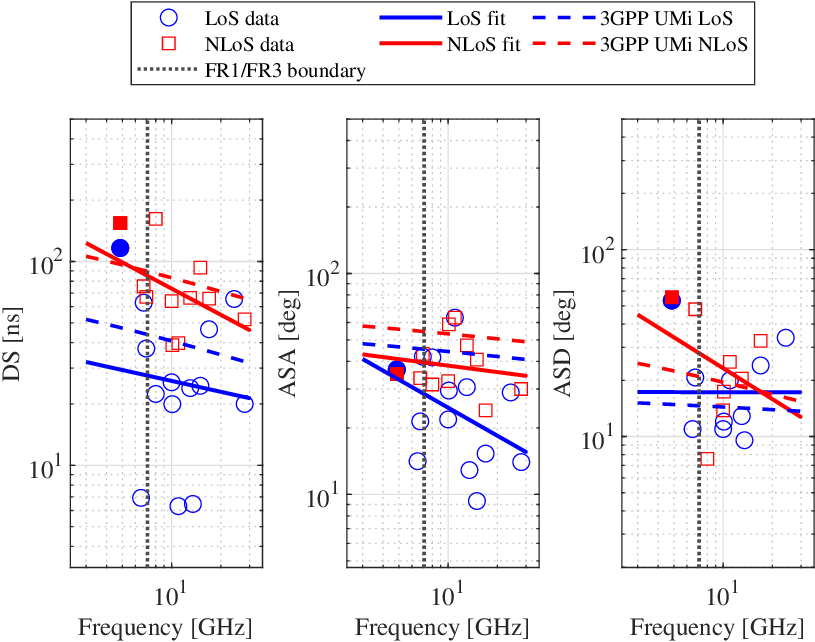}}

\caption{Literature-referenced cross-band trends for DS, ASA, and ASD. (The markers denote the anchor data points used in the analysis, with blue circles for LoS and red squares for NLoS; filled markers correspond to the $4.85$~GHz measurements from this work). The solid lines show the fitted indicative trends, while the dashed lines represent the corresponding 3GPP reference models. The vertical dotted line indicates the ($7.125$~GHz) boundary between FR1 and FR3 bands.
\label{fig:FC}}
\end{figure*}

\section{Literature-Referenced Cross-Band LSP Trends}

\subsection{Methodology}
To place the measured $4.85$~GHz LSPs in a broader yet bounded cross-band context, we performed a literature-referenced trend analysis using local route-wise mean values, together with selected scenario-matched literature anchors spanning approximately $4$--$28$~GHz. This range is chosen to focus on the Upper FR1/FR3 transition near the $7.125$~GHz boundary. The notation does not imply inclusion of FR2 or the broader mmWave regime. Although LSP values at higher mmWave frequencies are available in the literature \cite{Cheng_BeamDomain}, they are intentionally excluded because mmWave propagation is governed by substantially different blockage, penetration loss, scattering, atmospheric loss, and beamforming characteristics. A unified treatment that includes mmWave frequencies would require a separate modeling framework \cite{Kim-QD_Modeling}, which is beyond the scope of the present paper.

For each LSP $X\in\{\DS,\ASA,\ASD\}$, the frequency dependence is represented in log-log form as
\begin{equation}
\log_{10} X(f) = a_{X} \log_{10} f + b_{X},
\label{eq:Fc_model}
\end{equation}
where $f$ is the carrier frequency in GHz, $a_X$ is the log-log slope, and $b_X$ is the intercept. Equation~\eqref{eq:Fc_model} is an empirical power-law trend model introduced in this work for anchor-supported cross-band comparison; it is not a direct ITU-R prescribed model for DS, ASA, or ASD, although logarithmic frequency dependence is commonly used in propagation modeling \cite{ITU-R_P.1411}.
An important limitation of this literature-referenced analysis is that the anchor points were obtained from independent measurement campaigns with different hardware configurations, bandwidths, delay resolutions, dynamic ranges, calibration procedures, MPC extraction methods, and post-processing thresholds. This is particularly critical for DS, because the measurement bandwidth directly determines the resolvable delay resolution and can affect the estimated RMS delay spread by merging or separating closely spaced MPCs. Therefore, the DS anchors compiled from campaigns with bandwidths ranging from approximately $100$~MHz to more than $1$~GHz should not be regarded as strictly equivalent observations of the same underlying quantity. 
The robust regression used in this work can reduce the influence of isolated outlying anchors; it cannot remove systematic bias caused by non-harmonized measurement bandwidths and processing chains. Consequently, the fitted DS trend should be interpreted only as a qualitative literature-referenced indication, not as a calibrated multi-band DS parameterization.
\begin{table*}[htbp]
\centering
\caption{Leave-one-out (LOO) sensitivity of the UMi DS cross-band fit using the UMi DS anchors listed in Table~\ref{tab:Lit_Anchors}.}
\label{tab:loo_umi_ds}
\renewcommand{\arraystretch}{1.35}
\begin{tabular}{cccccc}
\hline
State & $N$ & $a_{\rm full}$ & LOO range of $a$ &
Largest $|\Delta a|$ anchor & Max. boundary change \\
\hline
LoS  & 14 & $-0.1775$ & [$-0.6981$,\,$0.0000$] &
16.95~GHz & $40.7$\% at 16.95~GHz \\
NLoS & 11 & $-0.4269$ & [$-0.5232$,\,$-0.1432$] &
4.85 GHz & $18.7$\% at 4.85~GHz \\
\bottomrule
\end{tabular}
\vspace{1mm}
\begin{flushleft}
\scriptsize
Here, $a$ denotes the slope coefficient in the log--log model $\log_{10}(X)=a\log_{10}(f)+b$. The full-anchor slopes are $a_{\rm unc}=-0.1775$ (LoS) and $a_{\rm unc}=-0.4269$ (NLoS); hence, the non-positive-slope constraint is inactive in both full fits (constrained/unconstrained). The LOO analysis is computed from the UMi DS anchors listed in Table~IV.
\end{flushleft}
\end{table*}

\begin{table*}[htbp]
    \centering
    \caption{Mean value ($\mu_{\log_{10}}$) models for DS, ASD, and ASA.}
    \label{tab:LSP_model_w_3gpp}
    \renewcommand{\arraystretch}{1.5}
    \begin{tabular}{ccccc}
    \toprule
    \textbf{Scenario} & \textbf{Cond.} & \textbf{Param.} & \textbf{3GPP \cite{TR138-901}}  & \textbf{This work} \\
    \midrule
    \multirow{6}{*}{UMa} 
      & \multirow{3}{*}{LoS}  
        & DS  & $-0.0963\log_{10}(f_c) - 6.955$ & $-0.25\log_{10}(f_c) - 7.02$ \\
      & & ASD & $-0.1114\log_{10}(f_c) + 1.06$ & $1.21$ \\
      & & ASA & $1.81$                         & $-0.15\log_{10}(f_c) + 1.76$ \\
      \cmidrule{2-5}
      & \multirow{3}{*}{NLoS} 
        & DS  & $-0.204\log_{10}(f_c) - 6.28$  & $-0.08\log_{10}(f_c) - 6.90$ \\
      & & ASD & $-0.1144\log_{10}(f_c) + 1.5$ & $-0.15\log_{10}(f_c) + 1.42$ \\
      & & ASA & $-0.27\log_{10}(f_c) + 2.08$  & $-0.03\log_{10}(f_c) + 1.74$ \\
    \midrule
    \multirow{6}{*}{UMi} 
      & \multirow{3}{*}{LoS}  
        & DS  & $-0.24\log_{10}(1+f_c) - 7.14$ & $-0.18\log_{10}(f_c) - 7.41$ \\
      & & ASD & $-0.05\log_{10}(1+f_c) + 1.21$ & 1.24 \\
      & & ASA & $-0.08\log_{10}(1+f_c) + 1.73$ & $-0.42\log_{10}(f_c) + 1.81$ \\
      \cmidrule{2-5}
      & \multirow{3}{*}{NLoS} 
        & DS  & $-0.24\log_{10}(1+f_c) - 6.83$ & $-0.43\log_{10}(f_c) - 6.71$ \\
      & & ASD & $-0.23\log_{10}(1+f_c) + 1.53$ & $-0.55\log_{10}(f_c) + 1.91$ \\
      & & ASA & $-0.08\log_{10}(1+f_c) + 1.81$ & $-0.10\log_{10}(f_c) + 1.68$ \\
    \bottomrule 
    \end{tabular}
    \vspace{1mm}
    \begin{flushleft}
    \footnotesize
    {Note: $f_c$ is the center frequency in GHz. For DS, the fitted intercepts are reported on the $\log_{10}(\mathrm{s})$ scale. The 3GPP UMi expressions use $\log_{10}(1+f_c)$, whereas the fitted models in this work use $\log_{10}(f_c)$.}
    \end{flushleft}
\end{table*}
The local route-wise mean values at $4.85$~GHz are first computed separately for each LSP under LoS/NLoS conditions. These measured means are then combined with the literature anchors listed in Table~\ref{tab:Lit_Anchors}, and all available points are fitted jointly. The anchors in Table~\ref{tab:Lit_Anchors} are not intended to constitute an exhaustive survey of all published LSP measurements. They are selected as scenario-matched UMa/UMi DS, ASA, and ASD reference points for a bounded trend analysis over the approximately $4$--$28$~GHz range. The upper limit of $28$~GHz is used only as the boundary of the considered literature-referenced range and should not be interpreted as extending the analysis to general FR2/mmWave propagation. Frequencies above this range are excluded because mmWave and sub-THz channels are subject to greater blockage, penetration loss, diffraction loss, atmospheric attenuation, antenna-array effects, and deployment-specific beamforming, which would require a separate modeling treatment \cite{Suzuki_SC}.
For each LSP, let $u_i=\log_{10} f_i$ and $v_i=\log_{10} X_i$. When $N\geq 3$ valid anchor points are available, the coefficients in \eqref{eq:Fc_model} are estimated by an adapted constrained robust M-fit,
\begin{equation}
(a_X,b_X)=\arg\min_{\substack{a\le 0\\ b\in\mathbb{R}}}
\sum_{i=1}^{N}\rho\!\left(\frac{r_i}{s_r}\right),
\label{eq:robustfit}
\end{equation}
where $r_i=v_i-(au_i+b)$ denotes the residual for anchor $i$, and $s_r=\operatorname{MAD}(r_i)/0.6745$ is the robust residual scale obtained from an initial unconstrained least-squares fit. The loss $\rho(\cdot)$ is taken as Tukey's bisquare,
\begin{equation}
\rho(r)=
\begin{cases}
1-\bigl(1-(r/k)^2\bigr)^3, & |r|\le k,\\[1ex]
1, & |r|>k,
\end{cases}
\label{eq:tukey}
\end{equation}
with $k=4.685$. The use of robust M-estimation, Tukey's bisquare weighting, and MAD-based scale normalization follows standard robust-regression practice \cite{HollandWelsch1977, BeatonTukey1974}. In contrast, the inequality constraint $a\le 0$ is introduced here as a physics-motivated monotonicity constraint to suppress non-physical positive slopes under sparse or heterogeneous anchor sets.
Because these results rely on a single well-calibrated $4.85$~GHz band and a limited set of heterogeneous literature anchors, the inferred cross-band slopes are interpreted as indicative rather than definitive, and they do not support firm conclusions or direct recommendations for 3GPP or ITU channel model parameterization.

\subsection{Fitting Results}
Figure~\ref{fig:FC} summarizes the literature-referenced cross-band trends for DS, ASA, and ASD in UMa and UMi under LoS and NLoS conditions. The local $4.85$~GHz anchors used in the figure are listed in Table~\ref{tab:Lit_Anchors}, Table~\ref{tab:loo_umi_ds} summarizes the leave-one-out (LOO) sensitivity of the UMi DS fit using the local DS anchors, and the fitted expressions are listed in Table~\ref{tab:LSP_model_w_3gpp}. These curves are intended only as anchor-supported trends over approximately $4$--$28$~GHz, providing a broader context for the measured $4.85$~GHz results and enabling comparison with the corresponding 3GPP reference trends.
The fitted behavior depends on the available anchors. Accordingly, the fitted expressions should be interpreted as literature-referenced, measurement-informed indications within the anchor-supported range, rather than as replacements for the corresponding 3GPP parameterizations or as validated scenario-wide frequency models.

\begin{itemize}
\item UMa: The fitted LoS DS trend decreases more noticeably with frequency, whereas the NLoS DS trend is nearly flat. The ASA and ASD fitted slopes remain conditional on heterogeneous literature measurements, bandwidths, and processing procedures. The UMa curves should be interpreted as bounded literature-referenced indications over approximately $4$--$28$~GHz, not as validated scenario-level models.
\item UMi: The fitted UMi trends indicate stronger frequency dependence than the corresponding 3GPP reference curves, particularly for NLoS DS and the angular spreads.  The LOO sensitivity in Table~\ref{tab:loo_umi_ds} for the LoS and NLoS fits use $14$ and $11$ valid anchors, respectively, including the local $4.85$~GHz anchor. For LoS DS, the full-anchor slope is $a=-0.1775$, and the LOO slopes range from $-0.6981$ to $0.0000$, with a maximum absolute slope change of $0.5206$ when the $16.95$~GHz anchor is omitted. This omission also gives the largest boundary-value change, increasing the fitted DS at $7.125$~GHz from $27.55$~ns to $38.76$~ns ($40.7\%$). For NLoS DS, the full-anchor slope is $a=-0.4269$, and the LOO slopes range from $-0.5232$ to $-0.1432$, with a maximum absolute slope change of $0.2837$ when the local $4.85$~GHz anchor is omitted. In that case, the fitted DS at $7.125$~GHz changes from $85.11$~ns to $69.21$~ns ($18.7\%$).

For both full-anchor fits, the corresponding unconstrained slopes are identical to the constrained values ($-0.1775$ for LoS and $-0.4269$ for NLoS); therefore, the non-positive-slope constraint is inactive in the full fits. Among the LOO fits, the constraint is active only for the LoS cases obtained by omitting the local $4.85$~GHz anchor and the $6.75$~GHz anchor. Their unconstrained slopes are $+0.2862$ and $+0.2012$, respectively, and the constrained estimates are consequently reported as $0.0000$. The constraint is inactive for all NLoS LOO fits. Thus, the fitted UMi DS slopes remain sensitive to the anchor set, particularly for LoS, and should be interpreted as literature-referenced cross-band indications rather than validated multi-band parameterizations.
\end{itemize}

The LOO analysis was reported for UMi DS because it has the largest number of valid anchors among the present LSP fits, making it the most meaningful case for assessing anchor sensitivity. It should be interpreted only as a sensitivity check, not as validation that removes bandwidth- or processing-dependent bias from the compiled DS anchors. The resulting trends, therefore, remain indicative rather than robust scenario-level parameterizations. Overall, this section provides only a bounded cross-band context for the $4.85$~GHz measurements, highlighting the continued need for consistent measurements and modeling data in the $4$--$28$~GHz range, especially in UMa.

\section{Conclusion}
This paper presented a measurement-based characterization of urban channels at $4.85$~GHz in three Yokohama UMa/UMi routes, including path loss (PL), delay spread (DS), azimuth spread of arrival and departure (ASA/ASD), a wideband snapshot-based $K$-factor proxy, and route-dependent spatial-consistency statistics. The results show that the measured large-scale behavior is strongly route- and geometry-dependent and should be interpreted in the context of the present $4.85$~GHz campaign.

To place these results in a broader context, the paper also includes a literature-referenced cross-band analysis for DS, ASA, and ASD over approximately $4$--$28$~GHz. This part is intended only as a bounded, measurement-informed view of the Upper FR1/FR3 transition region near $7.125$~GHz. Because it relies on a single in-house measurement band together with a heterogeneous set of literature, it should not be interpreted as a definitive multi-band model or as a basis for direct recommendations to 3GPP or ITU. The UMi DS leave-one-out analysis further showed that the fitted cross-band DS slopes are sensitive to individual anchors, reinforcing the interpretation of these curves as contextual trends rather than validated multi-band models.

Although the results are not presented as a definitive generalized model, they are useful to the community as a documented measurement reference for Upper FR1 urban propagation. The paper reports not only agreement with existing references but also route-dependent deviations, spatial-consistency behavior, and LSP statistics that can support future model calibration, site-specific prediction studies, and comparison with emerging FR3 datasets.

\end{document}